\documentclass[useAMS,usenatbib]{mn2e}
\usepackage{multicol}
\usepackage{psfig, epsf, epsfig}
\usepackage{amsmath}
\usepackage{pifont}% http://ctan.org/pkg/pifont
\newcommand{\cmark}{\ding{51}}%
\newcommand{\xmark}{\ding{55}}%
\title[Dust processes]{Photoelectric heating effects on the evolution of luminous disk galaxies}
\author[O. Osman, K. Bekki, and L. Cortese]
{Omima Osman ${}^{1,}$ ${}^2$\thanks{E-mail:
omima.osman@icrar.org},
Kenji Bekki${}^1$, and Luca Cortese${}^1$\\ 
${}^1$ICRAR M468
The University of Western Australia
35 Stirling Hwy, Crawley
Western Australia 6009, Australia\\
${}^2$
University of Khartoum - Department of Physics.
Al-Gamaa Ave, Khartoum 11115, Sudan.
P.O.Box 321}

\begin{document}

\date{Accepted, Received 2005 February 20; in original form }

\pagerange{\pageref{firstpage}--\pageref{lastpage}} \pubyear{2005}

\maketitle

\label{firstpage}

\begin{abstract}
Photoelectric heating (PEH) influences the temperature and density of the interstellar medium (ISM), and potentially also affecting star formation. PEH is expected to have a stronger effect on massive galaxies, as they host larger dust reservoirs compared to dwarf systems.  Accordingly, in this paper, we study PEH effects in Milky Way--like galaxies using smoothed particle hydrodynamics (SPH) code which self--consistently implements the evolution of the gas, dust, and interstellar radiation field (ISRF). Dust evolution includes dust formation by stars, destruction by SNe, and growth in dense media. We find that PEH suppresses star formation due to the excess heating that reduces the ISM density. This suppression is seen across the entire range of gas fractions, star formation recipes, dust models, and PEH efficiencies investigated by our code. The suppression ranges from negligible values to approximately a factor of five depending on the specific implementation. Galaxy models having higher gas fraction experience higher star formation suppression. The adopted dust model also alters the extent of star formation suppression. Moreover, when PEH is switched on, galaxy models show higher gas outflow rates and have higher loading factors indicative of enhanced SNe feedback. In gas--rich models (i.e. a gas fraction of 0.5), we also find that PEH suppresses the formation of disk clumps via violent disk instabilities, and thus suppresses bulge formation via clumps migration to the central regions.

\end{abstract}


\section{Introduction}
For several decades researchers have been studying the importance of the photoelectric heating (PEH) of the gas by dust
in the thermodynamical balance of the ISM (Watson 1972; Draine 1978; Bakes \& Tielens 1994; Wolfire et al. 2003;
Weingartner \& Draine 2001; Hill et al. 2018). Electrons ejected from dust grain surfaces by far--ultraviolet (FUV) photons are primarily responsible for the heating of the cold neutral medium (CNM) and diffuse atomic hydrogen (HI) regions (Wolfire et al. 1995; Ingalls, Reach, \& Bania 2002). Those electrons store their kinetic energy in the gas through collision with the different chemical species therein. PEH is very efficient in coupling FUV radiation ($6-13.6$ $eV$) to the gas, and it is mainly caused by the smallest size dust grains, i.e. Polycyclic Aromatic Hydrocarbons (PAH, Bakes \& Tielens 1994; Okada et al. 2013). The rate and efficiency ($F_e$) of the PEH  depend on the detailed ISM and dust properties such as the interstellar radiation field (ISRF) strength, ISM density and temperature, electron density, and grain sizes (see models proposed by, e.g.  Bakes \& Tielens 1994 and Weingartner \& Draine 2001).
 
Furthermore, PEH influences the dynamical evolution of galaxies through suppressing star formation as a result of the drop in gas density. However, the importance of PEH in regulating star formation is still arguable compared to the other feedback mechanisms such as SNe. For instance, using numerical simulation, Forbes et al. (2016)  investigated SNe feedback and PEH in dwarf galaxies and concluded that PEH is the dominant process in regulating star formation. On the other hand, Hu et al. (2017) found the contrary and reported that PEH is unable to create outflows and produce the multiphase structure of the ISM, unlike SNe. Hu et al. (2017) rigorously justified the discrepancy between their results and those of Forbes et al. (2016) by attributing it mainly to inconsistent cooling implementation by  Forbes et al. (2016).  For Milky Way--like galaxies, Tasker (2011) who studied the influence of star formation and PEH on the lifetime of giant molecular clouds (GMCs) found that PEH changes the ISM structure and suppresses the initial star formation (in the period between 50 and 200 Myrs).  After 200 Myrs, the GMCs where PEH is switched on have higher star formation rate (SFR) since they return enough gas to continue forming stars. It should be noted that Tasker's (2011) study did not include SNe feedback. Bekki's (2015) study on disk galaxies, which involved a sophisticated implementation of the dust hydrodynamical interaction with the rest of the galaxy components, also predicted SFR suppression when PEH is included in simulations.

PEH is commonly studied in the context of feedback in dwarf galaxies (e.g. Forbes et al. 2016; Hu et al. 2017; Emerick et al. 2019). This is justified by their small size and mass, which allow for high--resolution simulations, and the fact that they are chemically young objects (i.e. simple ISM to model). However, dwarf galaxies, have lower dust content compared to massive galaxies (Remy--Ruyer et al. 2014; Grossi et al. 2015) as the bulk of their stars (initial source of dust) formed at later epoch compared to massive galaxies (Cowie et al. 1996; Gavazzi et al. 1996). Moreover, dust growth in the ISM of dwarf galaxies requires a more extended time period to overcome dust production by stars (the critical metallicity to set on dust growth is low, but the galaxy takes a long time to reach it, Inoue 2011; Asano et al. 2013). Zhukovska (2014) found that dust growth in dwarf galaxies becomes important in a timescale of 0.1 to 1 Gyr. 

We expect PEH to have a significant influence on the evolution of more massive, luminous disk galaxies since they efficiently form dust through star formation and dust growth (see, e.g. Dwek 1998; Hirashita 2013; Aoyama et al. 2017 for discussions on the dust processing in the ISM). In particular, the PEH rate is directly proportional to the dust--to--gas ratio (Eq. 5 section 2.2.2 of this paper, Bakes \& Tielens 1994; Wolfire et al. 2003; Bergin et al. 2004) which has higher values in massive galaxies. Dust and gas are closely linked, and they interact through several processes that influence both (e.g. PEH, dust growth and destruction, molecular hydrogen formation, Hirashita 2000; Nozawa, Kozasa \& Habe 2006; Wakelam et al. 2017). Thus, we expect PEH also to influence the spatial distribution of the ISM components and regulate dust evolution. These two last points can not be resolved assuming a constant dust--to--gas ratio since dust evolution (varying dust--to--metal ratio) is proven to be relevant (e.g. Dwek \& Scalo 1980; Hirashita 2000;  Asano et al. 2013; Bekki 2015; Chiang et al. 2018).

Thus the purpose of this paper is to investigate the role of PEH in suppressing star formation in Milky way--like galaxies. We also study the dependence of this suppression on the gas fraction,  the PEH efficiency, and the galaxy mass. Moreover, the paper investigates how PEH  self--regulates dust evolution, gas and H$_2$ consumption, and their spatial distribution (the ISM structure). To accurately model PEH, one needs to simultaneously model the evolution of the gas, dust, and ISRF (Table 1 shows the dust physics included in the present models). Although several theoretical studies succeeded in modelling the time and space varying ISRF (e.g. Forbes et al. 2016; Hu et al. 2017; Emerick et al. 2019) using adaptive mesh refinement hydrodynamical codes and smoothed particle hydrodynamical (SPH) simulations, none of them modelled (i) dust formation and evolution, (ii) formation of H$_2$ on dust grains, and (iii) self--shielding of H$_2$ effects, all of which could be important for the ISM of luminous disk galaxies (Osman et al. 2020).

\vspace{0.3cm}
\begin{table}
\centering
\begin{minipage}{75mm}
\caption{Dust physics included in the models.}
\begin{tabular}{ccc}
\hline
{Dust physics}
& {This work}
& {Reference\footnote{1 Bekki (2013), 2 Aoyama et al. (2017), 3 Bekki (2015), 4 Forbes et al. (2016), 5 Hu et at. (2017), 6 Hirashita \& Inoue (2019), 7 Rollig et al. (2006, they derived analytical formula for electron density), 8 Hirashita \& Aoyama (2019), 9 Hou et al. (2017), 10 McKinnon et al. (2018), 11 Zhukovska et al. (2016), 12 Osman et al. (2020), 13 Nozawa, Kozasa \& Habe (2006), 14 Bakes \& Tielens (1994).}}\\
\hline
Dust formation & \cmark & 1, 2 \\
Dust destruction & \cmark & 1, 2, 3 \\
Dust growth & \cmark & 1, 2, 3 \\
Dust shattering and coagulation & \xmark & 2\\
Photoelectric heating & \cmark & 3, 4, 5 \\
Radiation Pressure on dust& \xmark  & 3, 6\\
Electron density of the ISM & \xmark  & 7\\
H$_2$ formation on dust grains & \cmark  & 1, 3\\
Grain size distribution & \xmark & 2, 8\\
Grain composition & \xmark & 9\\
Dust extinction & \cmark & 10, 3\\
Varying dust sticking coefficient & \xmark & 11, 12\\
Varying destruction efficiency & \xmark & 13\\
Dust grains dynamics & \xmark & 10\\
Dust--corrected cooling & \cmark & 1, 3\\
Grain charge distribution & \xmark & 14\\
\hline
\end{tabular}
\end{minipage}
\end{table}

Here we present a SPH simulation code that self--consistently models the evolution of the gas, dust, and ISRF. However, our code is still limited in its ability to model the PEH process adequately since it does not resolve the electron density in the ISM. The electron density is one of the quantities that determine $F_e$ along with the gas temperature and radiation field. For this reason, we are forced to treat $F_e$ as a parameter that ranges from 0.003 to 0.05 (Bakes \& Tielens 1994). From an observational point of view, the ratio between [CII] 158 $\micron$ line and the total infrared emission or between [CII] 158 $\micron$ + [OI] 63 $\micron$ and PAH emission is used as an estimate of the PEH efficiency (Croxall et al. 2012; Beirao et al. 2012; Herrera--Camus et al. 2018). Observations of NGC 1097 and NGC 4559 galaxies show a PEH efficiency of up to 0.06 (Croxall et al. 2012). Rollig et al. (2006) derived an analytical formula for electron density in photon--dominated regions using KOSMA$-\tau$ model (Storzer et al. 1996).

The rest of this paper is organized as follows: sections 2 and 3 describe the simulations set up and the main results of the study, respectively.  Discussion on the results is given in section 4, while section 5 contains the conclusions and summary of the paper.

\section{The model}

The SPH chemodynamical model used in the present study is a modified version of the model presented in Bekki (2013) and Osman et al. (2020) (hereafter B13 and O20). In this updated version, we advantage from the detailed modelling of the dust evolution, and the time and space varying ISRF presented in O20 to introduce PEH into the model self--consistently.  Hence, we briefly describe the model and refer the reader to B13 and O20 for further details. This study models an isolated Milky Way--like (MW--like) disk with NFW (Navarro, Frenk \& White 1996) dark matter halo density profile. Initially, the gaseous halo is in hydrostatic equilibrium and has a NFW density profile. Tables 2 and 3 show some of the underlying parameters and physics included in the models.

\subsection{Chemical enrichment and dust model}
The chemical abundance of eleven elements such as He, C, N, O, Mg and Ca is followed in time after their ejection by SNIa, SNII, and AGB stars (B13; Bekki 2015). Metals ejected are equally distributed among the neighbouring gas particles and non--instantaneous recycling is assumed (e.g., Bekki \& Shioya 1998). A fraction of the ejected metals condenses to form dust grains or accretes onto dust grain surfaces in the ISM (B13). We adopt the dust model proposed by Dwek (1998) here. The model accounts for: (a) dust formation by AGB stars, SNIa, and SNII, (b) dust growth by accretion of the ISM gas--phase metals, (c) dust destruction by SNe blasts, and (d) formation of polycyclic aromatic hydrocarbons (PAHs). The dependence of the dust growth and destruction processes on the ISM properties (temperature, density, and metallicity) is accounted for, i.e. the timescale of those processes is not constant throughout the galaxy (O20). H$_2$ formation on dust grains is also self--consistently implemented with the dust evolution (B13).

\subsection{Star formation and feedback}
For star formation, we mainly apply a H--dependent star formation recipe (SFR depends on the total density of the gas), however, a H$_2$--dependent recipe (SFR depends on the H$_2$ density, see B13) is used for comparison with the H--dependent recipe results in one model. In H--dependent recipe, a gas particle is transformed into a stellar particle with Salpeter IMF if: the local density exceeds a threshold density  ($\rho_{th}$ = 1 cm$^{-3}$ in the present study), the local velocity field is consistent with gravitational collapse (i.e. div $v < 0$), and the sound crossing timescale is longer than the local dynamical timescale. When the formed stars explode as SNe, the total energy injected into the ISM  ($E_{\rm sn}$ = $10^{51}$ erg per SNa) is divided into 90\% thermal feedback and 10\% kinematic feedback consistent with the numerical simulations by Thornton et al. (1998). This energy is distributed equally among the neighbouring particles (Bekki et al. 2013).The radiative cooling is implemented using the cooling curve by Rosen \& Bregman (1995) for 100 $\leq$ T $<$ $10^4$ K and the MAPPING III code for T $\geq$ $10^4$ K (Sutherland \& Dopita 1993). The gas also cools down beyond 100 K via H$_2$ cooling to a floor of 10 K. The implementation of the PEH associated with the dust is described in the following subsection.

\begin{table}
\centering
\begin{minipage}{70mm}
\caption{Description of the basic parameters values
for the MW--like models.}
\begin{tabular}{cc}
\hline
{Physical properties}
& {Parameter values}\\
\hline
{Total Mass \footnote{$M_{\rm h}=M_{\rm dm}+M_{\rm g}$, where
$M_{\rm dm}$ and $M_{\rm g}$ are the total masses of dark matter halo
and gas in a galaxy, respectively.}}
 & $M_{\rm h}=10^{12} {\rm M}_{\odot}$  \\
{Structure \footnote{ For the structure of the dark matter halo NFW profile with a virial radius ($r_{\rm vir}$) and a $c$--parameter is adopted.}}
 & $r_{\rm vir}=245$ kpc,  $c=10$  \\
Initial H$_2$ fraction & 0.01     \\
Initial metallicity   &   ${\rm [Fe/H]_0}= 0.30$ dex \\
 Metallicity gradient & $-0.04$ dex/kpc \\
Initial dust/metal ratio  & 0.4  \\
{SF \footnote{$\rho_{\rm th}$ is the threshold gas density for star formation. The interstellar radiation field (ISRF) is included in the estimation of  ${\rm H_2}$ mass fraction in this model.}}
 & ISRF,  $\rho_{\rm th}=1$ cm$^{-3}$ \\
IMF & Salpeter ($\alpha=2.35$) \\
Softening length  \footnote{$\epsilon_{\rm dm}$ and $\epsilon_{\rm g}$ are for the dark matter and gas, respectively.}  & $\epsilon_{\rm dm}=935$ pc, $\epsilon_{\rm g}=94$ pc \\
Gas mass resolution & {$m_{\rm g}\sim3\times10^4 {\rm M}_{\odot}$} \\
\hline
\end{tabular}
\vspace{-0.15cm}
\end{minipage}
\end{table}

%%%%%% TABLE1'
\begin{table}
\centering
\caption{The grid of the models analyzed for this study. f$_g$, f$_b$, $F_e$, and SF recipe are the gas fraction, baryonic fraction, PEH efficiency, and the star formation recipe dependent on either the total gas (H) or molecular gas (H$_2$) densities. Dust evolution (due to formation by stars, growth in dense media, and destruction by SNe) is switched on in the model if indicated \cmark. M14 and M15 are models of interacting pair of galaxies. M19 and M20 are models with baryon fraction 50\% less than the rest of the models (i.e., f$_b$ = 0.03).}

\begin{tabular}{clllcc}
\hline
{Model ID}  
 &  f$_g$
  &  f$_b$
 &  $F_e$
 & SF recipe
 &  Dust evolution\\
\hline
M1 &  0.1 & 0.06 & 0.05 & H & \cmark \\
M2 &  0.1 &  0.06 & 0.0 & H & \cmark \\
M3 &  0.5 &  0.06 &  0.05 & H & \cmark \\
M4 &  0.5 &  0.06 &  0.0 & H & \cmark \\
M5 &  0.3 &  0.06 &  0.05 & H & \cmark \\
M6 &  0.3 &  0.06 &  0.0 & H & \cmark \\
M7 &  0.03 &  0.06 &  0.05 & H& \cmark \\
M8 &  0.03 &  0.06 &  0.0 & H& \cmark \\
M9 &  0.1 &  0.06 &  0.03 & H & \cmark \\
M10 &  0.1 &  0.06 &  0.01 & H & \cmark \\
M11 &  0.1 &  0.06 &  0.003 & H & \cmark \\
M12 &  0.1 &  0.06 &  0.05 & H$_2$ & \cmark \\
M13 &  0.1 &  0.06 &  0.05 & H & \xmark \\
M14 &  0.1 &  0.06 &  0.05 & H & \cmark \\
M15 &  0.1 &  0.06 &  0.0 & H & \cmark \\
M16 &  0.5 &  0.06 &  0.03 & H & \cmark \\
M17 &  0.5 &  0.06 &  0.01 & H & \cmark \\
M18 &  0.5 &  0.06 &  0.003 & H & \cmark \\
M19 &  0.5 &  0.03 &  0.05 & H & \cmark \\
M20 &  0.5 &  0.03 &  0.0 & H & \cmark \\

\hline

\end{tabular}
\vspace{-0.15cm}
\label{table3}
\end{table} 

\subsubsection{Photoelectric heating}
We implement the same PEH model implemented in Bekki (2015) where the analytic formula for the photoelectric heating rate (n$\Gamma_{pe}$) proposed by Bakes \& Tielens (1994) is adopted:

\begin{equation}
n\Gamma_{pe} = \begin{array}{rcl} \beta F_e nG_0  & \rm{erg cm^{-3} s^{-1}} \end{array}
\end{equation}
where $\beta$ = $1\times 10^{-24}$, n, $F_e$, and $G_0$ are the number density of the ISM, PEH efficiency, and the intensity of the FUV field in units of Habing (1968), respectively.

The PEH rate and efficiency ($F_e$) depend on the underlying ISM and dust properties such as the ISRF strength, ISM density and temperature, electron density, and grain sizes (Bakes \& Tielens 1994; Weingartner \& Draine 2001). To estimate the time and space varying $G_0$ in the present model, we follow the following steps:

1- \textit{estimate the FUV radiation strength}. For each stellar particle, the stellar population synthesis codes by Bruzual \& Charlot (2003) are used to estimate its SED according to its age and metallicity, and hence estimate the strength of the FUV part of the ISRF. Not all the FUV radiation is used in PEH, part of which is exhausted by dust extinction. Thus the flux at a wavelength $\lambda$ in the FUV part from the $i$th stellar particle around the $j$th gas particle is given by the following equation according to the screen model:

\begin{equation}
f_{\lambda,i} = f_{\lambda,0,i} e^{-\tau_{\lambda,j}r_{i,j}/h_{j}},
\end{equation}
where $f_{\lambda, 0, i}$, $\tau_{\lambda, j}$ , $r_{i,j}$, and $h_{j}$ are the original flux of the stellar particle, FUV optical depth, distance between the gas and the stellar particles, and SPH smoothing length of the gas particle.

2- \textit{estimate the fraction of light absorbed by dust}, this fraction is calculated using the following equation:

\begin{equation}
F_{ex,i,j} = \int_{0}^{1} e^{-\tau_{\lambda, j}r_{i,j}} dr,
\end{equation}
where for the adopted SPH kernel (equals  0 at $h_j$), the integration in this equation ranges between 0 and 1. To estimate $\tau_{\lambda, j}$ for each gas particle, first the optical dust extinction ($A_{V,j}$) is estimated using the gas column density and the dust--to--gas ratio. Then, $A_{FUV,j}$ is estimated using the Calzetti's extinction law which relates $A_{V,j}$ and $A_{FUV,j}$ (i.e. Eq. 4 in Calzetti et al. 2000), and thus $\tau_{\lambda, j}$ ($\tau_{\lambda} \approx 0.921A_{\lambda}$). The gas column density is estimated using the 3D hydrogen density and $h_j$, i.e. $\rho_j(H)h_j$. Hence $G_{0,i,j}$ is given by the following equation:

\begin{equation}
G_{0,i,j} = F_{ex,i,j}g_{0,i,j},
\end{equation}
where $g_{0,i,j}$ is $G_{0,i,j}$ assuming no dust extinction.

Estimating the ISM electron density involves following a network of chemical reactions of the abundant elements in the ISM (Rollig et al. 2006) as well as the grain size and charge distributions (Bakes \& Tielens 1994; Weingartner \& Draine 2001; Wolfire et al. 2003) which are outside the scope of this study. Accordingly, we are forced to treat the $F_e$ as a parameter that ranges from 0.003 to 0.05 (Bakes \& Tielens 1994). 

\begin{figure*}
\includegraphics[width=1.\linewidth]{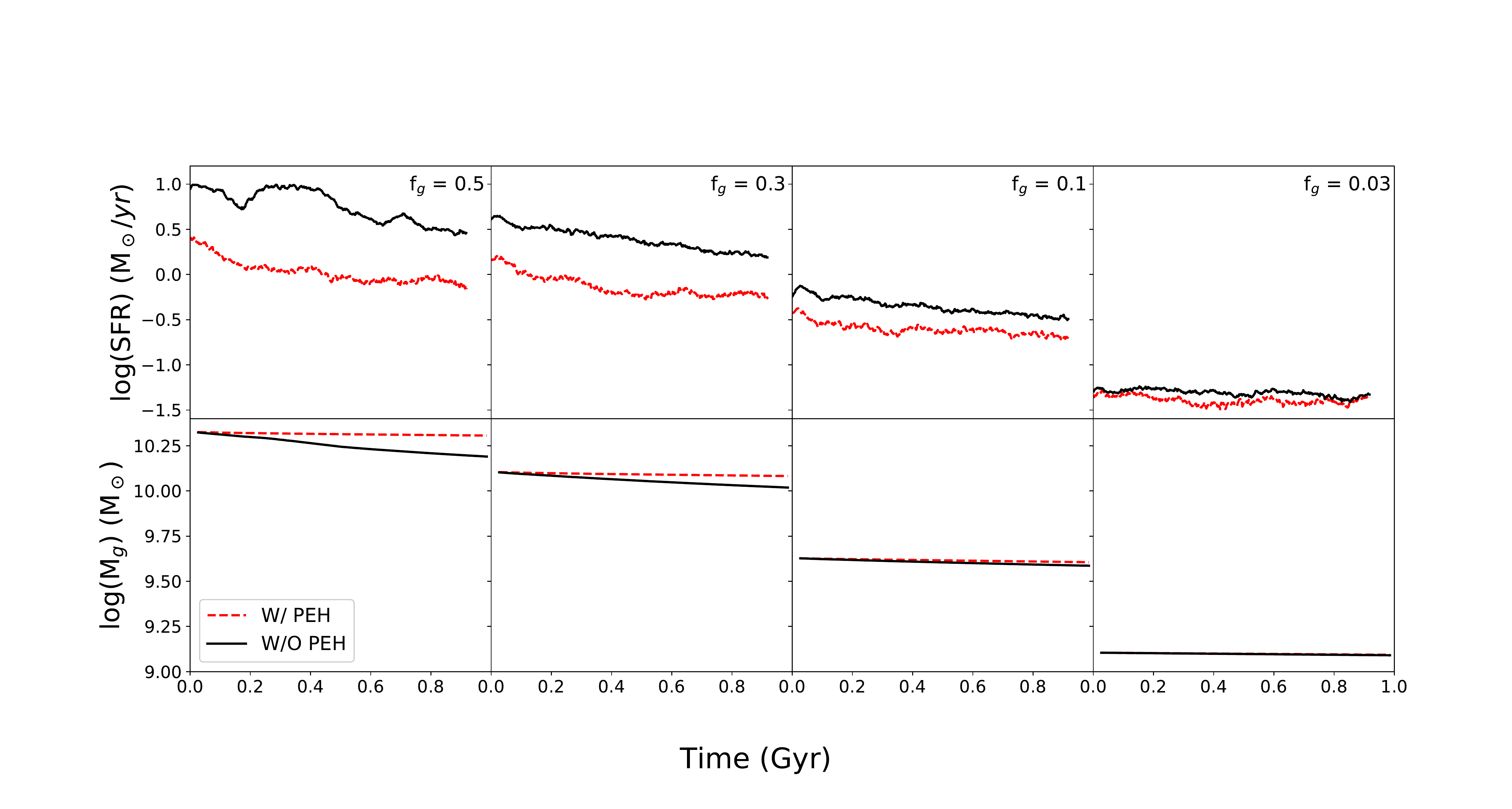}\par 
\vspace{-0.6cm}
\caption{The time evolution of the SFR (upper row) and the total gas mass (bottom row) in models M3, M4 (f$_g$ (gas fraction) = 0.5), M5, M6 (f$_g$ = 0.3), M1, M2 (f$_g$ = 0.1), and M7, M8 (f$_g$ = 0.03). The red dashed and black solid lines represent models with $F_e$ = 0.05 (W/ PEH) and $F_e$ = 0.0 (W/O PEH), respectively.}
\label{fig1}
\end{figure*} 

\begin{figure}
\includegraphics[height=6.cm,width=8.cm]{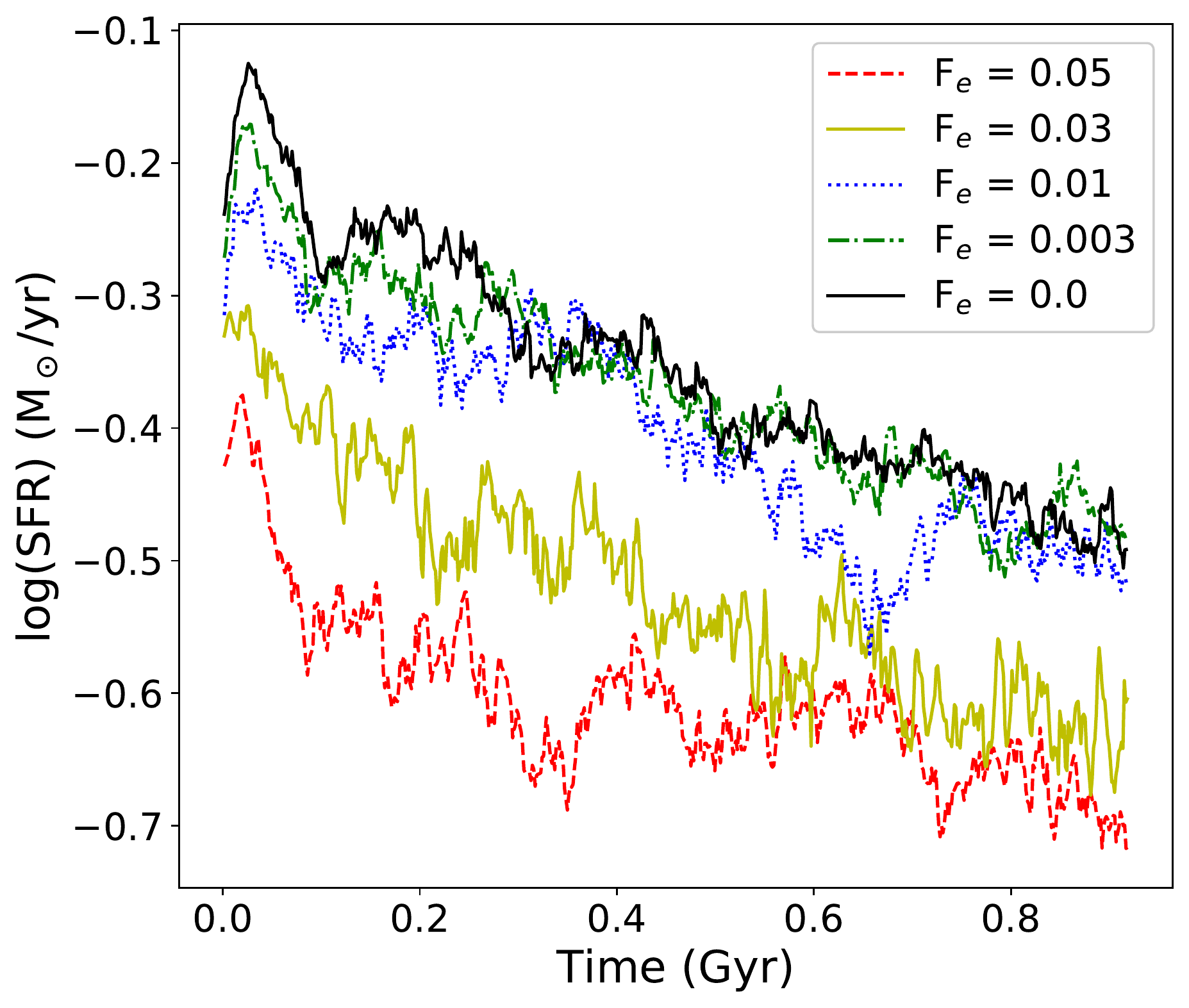}\par 
\vspace{-0.2cm}
\caption{The time evolution of the SFR in models M1, M9, M10, M11, and M2 with $F_e$ = 0.05, 0.03, 0.01, 0.003, and 0.0 and f$_g$ = 0.1, respectively.}
\label{fig2}
\end{figure}
\subsubsection{Photoelectric heating in previous studies}
Several recent hydrodynamical simulations implemented PEH in the study of the ISM cloud evolution, feedback in dwarf galaxies, and evolution of disk galaxies. In the following, we review the PEH implementation in a few of these studies. Most of these studies used the analytic formula proposed by Bakes \& Tielens (1994) in one form or another.

In dwarf galaxies,  Hu et al. (2017) and Emerick et al. (2019) adopted the formula in Eq. 2, following Bakes \& Tielens (1994), Wolfire et al. (2003), and Bergin et al. (2004). Both studies were able to track free electrons in their models.  However, Emerick et al. (2019) applied $F_e$ that varies with the gas density in the form $F_e$ = 0.0148n$^{0.235}$ (because their chemical network does not track electrons from all essential elements), and they used Remy--Ruyer et al. (2014) fit to estimate the dust--to--gas ratio. Hu et al. (2017) found PEH to be a subdominant feedback mechanism in dwarf galaxies, while  Emerick et al. (2019)  reported that the multichannel feedback (including PEH) in their models resulted in a realistic evolution of a dwarf galaxy that is consistent with observations in terms of star formation and metallicity of outflows. Forbes et al. (2016) used a recipe in which the PEH rate is directly proportional to the flux of the FUV photons and the density of metals ($Zn_H$) with the approximation that temperature and electron density have negligible effects in cold/dense media. 

\begin{equation}
\Gamma_{pe} = \begin{array}{rcl} 1.3\times \beta F_e nG_{eff}D & \rm{erg cm^{-3} s^{-1}} \end{array}
\end{equation}
where $G_{eff} =G_0e^{(-1.33\times10^{-21}DN_{H, tot})}$ is the attenuated radiation field strength in units of Habing and D is the dust--to--gas ratio. Although Eqs 1 and 2 appear similar, the underlying physics of the dust and ISM is different.

In disk galaxies, Butler et al. (2017) used the same formula used by Forbes et al. (2016) for the PEH rate in their kpc study on how stellar feedback regulates SFR. The authors concluded that including only SNe feedback results in similar SFRs to the SFRs in their models where PEH and ionization due to the FUV and extreme ultraviolet (EUV) radiation, respectively, are included as well. However, the ISM has very different temperature and chemical states, and the young stars have a different distribution.  Bekki (2015) implementation is the same as the implementation we are presenting here with $F_e$ = 0.003. Tasker (2011) studied the effect of the PEH and star formation on the GMCs formation and evolution using Eq. 3 for the PEH rate following Wolfire et al. (2003). 
\begin{equation}
 \Gamma_{pe} = \beta F_eG_{0}
  \begin{cases} 
   \exp{-(R-R_0)/H_R} & ergs^{-1} r \geq 4  kpc \\
   \exp{-(4-R_0)/H_R} &  ergs^{-1} r < 4  kpc
  \end{cases}
\end{equation}
where $F_e$ = 0.05, $H_R$, and $R_0$ are the scale length (= 4.1 kpc, Wolfire et al. 2003) and the radial scale length at 8 kpc, respectively. Other implementations can also be found in Choi et al. (2017) and Hill et al. (2018). In all of those studies, dust evolution was not explicitly followed.

\section{Results}

\subsection{PEH effects on SFRs}
Fig. \ref{fig1} shows the time evolution of the SFR (upper row) and the total gas mass (bottom row) in models M3, M4 (f$_g$ (gas fraction) = 0.5), M5, M6 (f$_g$ = 0.3), M1, M2 (f$_g$ = 0.1), and M7, M8 (f$_g$ = 0.03). The red dashed, and black solid lines represent models with $F_e$ = 0.05 and 0.0, respectively. When $F_e$ is greater than zero, the PEH is switched on (W/ PEH) in the model with efficiency $F_e$, and it is switched off (W/O PEH) when $F_e$ = 0.0. Table 3 gives $F_e$ values for the different models.

The diffuse heating caused by the PEH results in an ISM with average temperature one order of magnitude higher in M1, M3, and M5 compared to M2, M4, and M6, and twice as hot in M7 compared to M8. Thus, the density of the ISM drops in models with PEH compared to models without PEH. The rise in temperature and decrease in density increase the stability of the gas particles in M1, M3, M5, and M7 against gravitational collapse to form stars (lower density and higher Jeans' mass) compared to the gas particles in M2, M4, M6, and M8 models . Accordingly, the SFR (upper row) and gas consumption (bottom row) in models with PEH are lower than in models without PEH. In the absence of late episodic or continuous gas accretion (isolated disk galaxy), the SFR gradually declines by 0.6 to 0.2 dex as the gas fraction decreases from 0.5 to 0.03, respectively. This decline is due to the gas consumption by star formation.

In addition to the SFRs suppression in models with PEH compared to models without PEH, the magnitude of suppression (how much SFR is reduced when PEH is switched on) increases with the gas fraction, which can also be seen in the gas evolution (bottom row). For instance, when the gas fraction is increased from 0.1 in M1 and M2 (middle--right column) to 0.5 in M3 and M4 (left column), the SFRs rose considerably, especially in the case where PEH is switched off. The average SFRs increased by 0.65 dex in M3 compared to M1 and by 1.14 dex in M4 compared to M2. Moreover, PEH suppressed star formation by 0.74 dex in M3 compared to M4, while a suppression of 0.26 dex occurred in M1 compared to M2.

\begin{figure}
\includegraphics[height=8cm,width=9.cm]{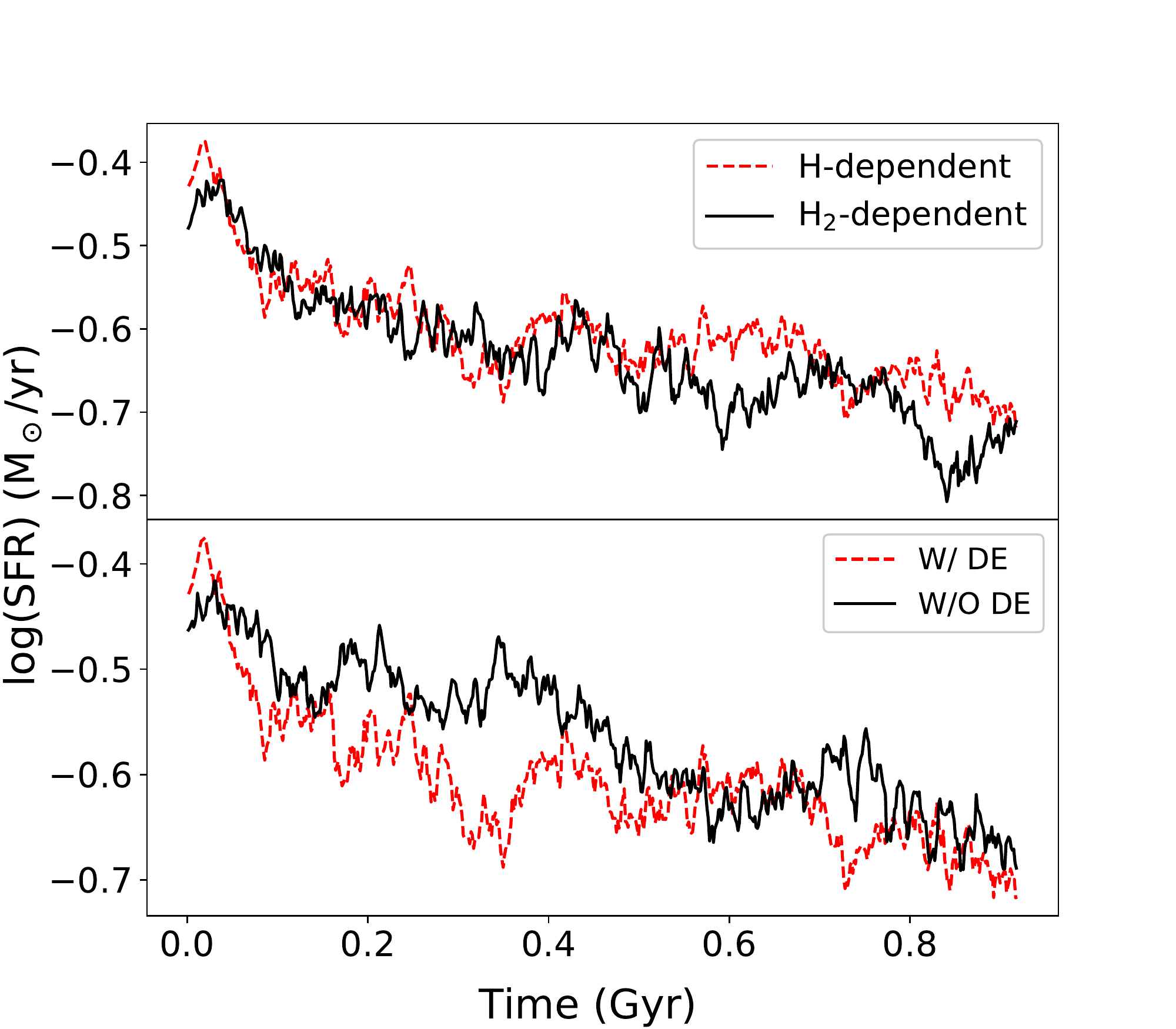}\par 
\vspace{-0.2cm}
\caption{The time evolution of the SFR in models M1 (red dashed, $F_e$ = 0.05) and M12 (black solid, $F_e$ = 0.05) with H and H$_2$ dependent SF recipes, respectively, top panel. The bottom panel shows the same in models M1 (red dashed, W/ DE) and M13 (black solid, W/O DE) with and without dust evolution, respectively.}
\label{fig3}
\end{figure} 

One of the models' limitations is the unresolved electron density which forced us to treat $F_e$ as a parameter that ranges from 0.05 to 0.003 (Bakes \& Tielens 1994). Fig. \ref{fig2} shows the time evolution of the SFR in models M1, M9, M10, M11, and M2 with $F_e$ = 0.05, 0.03, 0.01, 0.003, and 0.0 and f$_g$ = 0.1, respectively. The suppression of the SFR, as in the case of gas fraction, is present in all models with PEH regardless of $F_e$ value, however, its magnitude depends on the actual value of $F_e$ adopted (increases with $F_e$ ). $F_e$ is not constant and varies according to the ISM conditions, hence adopting one value for $F_e$ throughout the galaxy results in error in estimating the mean SFR corresponding to a standard deviation of up to 0.1 dex. The same argument holds for models M16, M17, and M18 with gas fraction 0.5, where the estimation of the SFR is associated with an error corresponding to 0.2 dex standard deviations.

The physics included in the models influences the resultant SFRs as well. In Fig. \ref{fig3} we examine the influence of the SF recipe and the dust model on the previous results. The top panel shows the time evolution of the SFR in models M1 (red dashed) and M12 (black solid) with H and H$_2$ dependent SF recipes, respectively. In both models, PEH is switched on ($F_e$ = 0.05). Adopting SFR that scales with the total gas density rather than with the H$_2$ density results in a slight overestimation of the SFR (0.02 dex on average). The density of the total gas is more often around the threshold density for the star formation, unlike H$_2$ density, which results in a reduction of the star formation. The two models share similar SFRs only when the gas is mostly molecular. The slightly lower SFR in M12 beyond 500 Myrs resulted in dust build--up which, in turn, suppressed the star formation further through PEH. With a slight delay, the build--up of dust enhanced H$_2$ abundance. In this case, where the ISM has significantly high molecular hydrogen fraction, adopting either of the star formation recipes results in a small difference. On the contrary, in the case of the model without PEH, the difference is more significant since it has a lower hydrogen fraction throughout the course of the evolution.

The bottom panel of Fig. \ref{fig3} shows the time evolution of the SFR in models M1 (red dashed, W/ DE) and M13 (black solid, W/O DE). Dust evolution due to formation by stars, destruction by SNe, and growth in dense media is implemented in M1. In M13, constant dust--to--gas ratio (D) is used instead. PEH is implemented in both models with $F_e$ = 0.05. Implementing constant D results in an overestimation of the SFR (0.04 dex on average) since D decreases steadily as the galaxy evolves and consumes its gas which suppresses the PEH effect. In the model with dust evolution, dust abundance increases gradually with time until it reaches a peak before declining at later epochs. The adopted dust model would not influence the SFR in models without PEH and with  H--dependent SF recipe since these models are hardly sensitive to how much dust is present in the ISM. This implies that models that implement constant D underestimate or overestimate PEH effect in suppressing SFRs depending on the adopted D.

\subsection{PEH effects on the gas and dust distributions}

\begin{figure*}
	\begin{multicols}{2}
    \includegraphics[height=5cm,width=8.5cm]{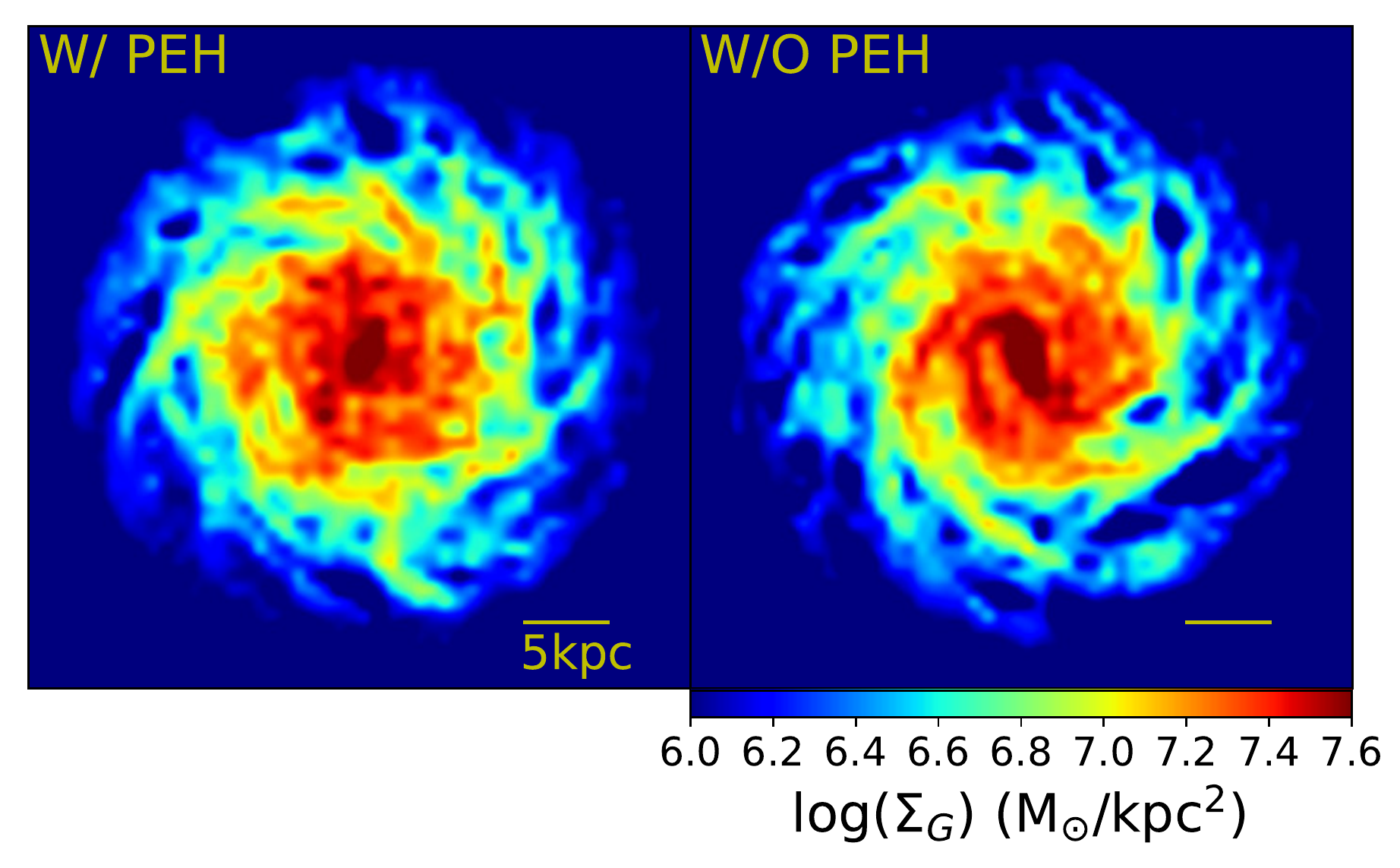}\par 
    \includegraphics[height=5cm,width=8.5cm]{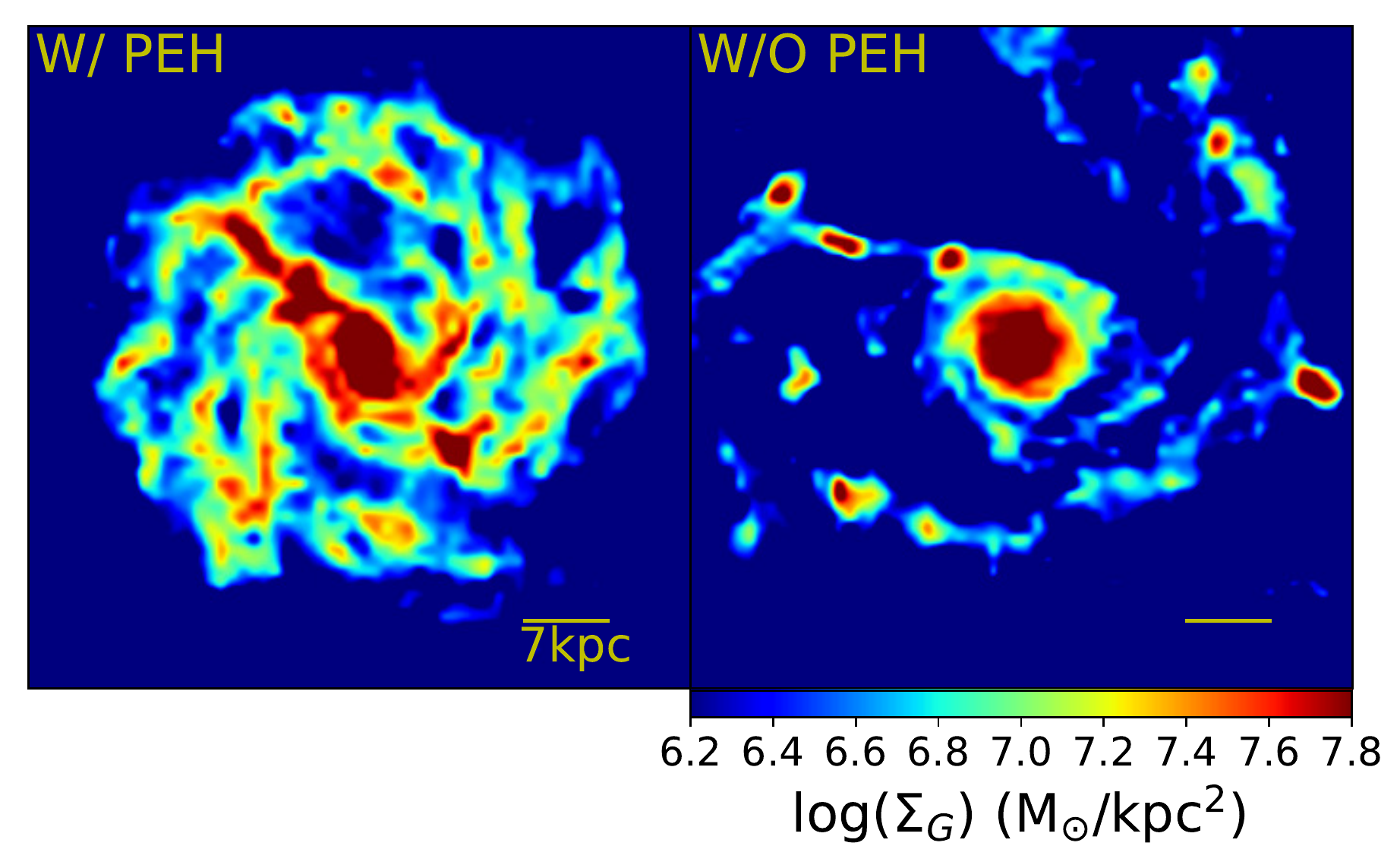}\par 
    \end{multicols}
    \begin{multicols}{2}
    \includegraphics[height=5cm,width=8.5cm]{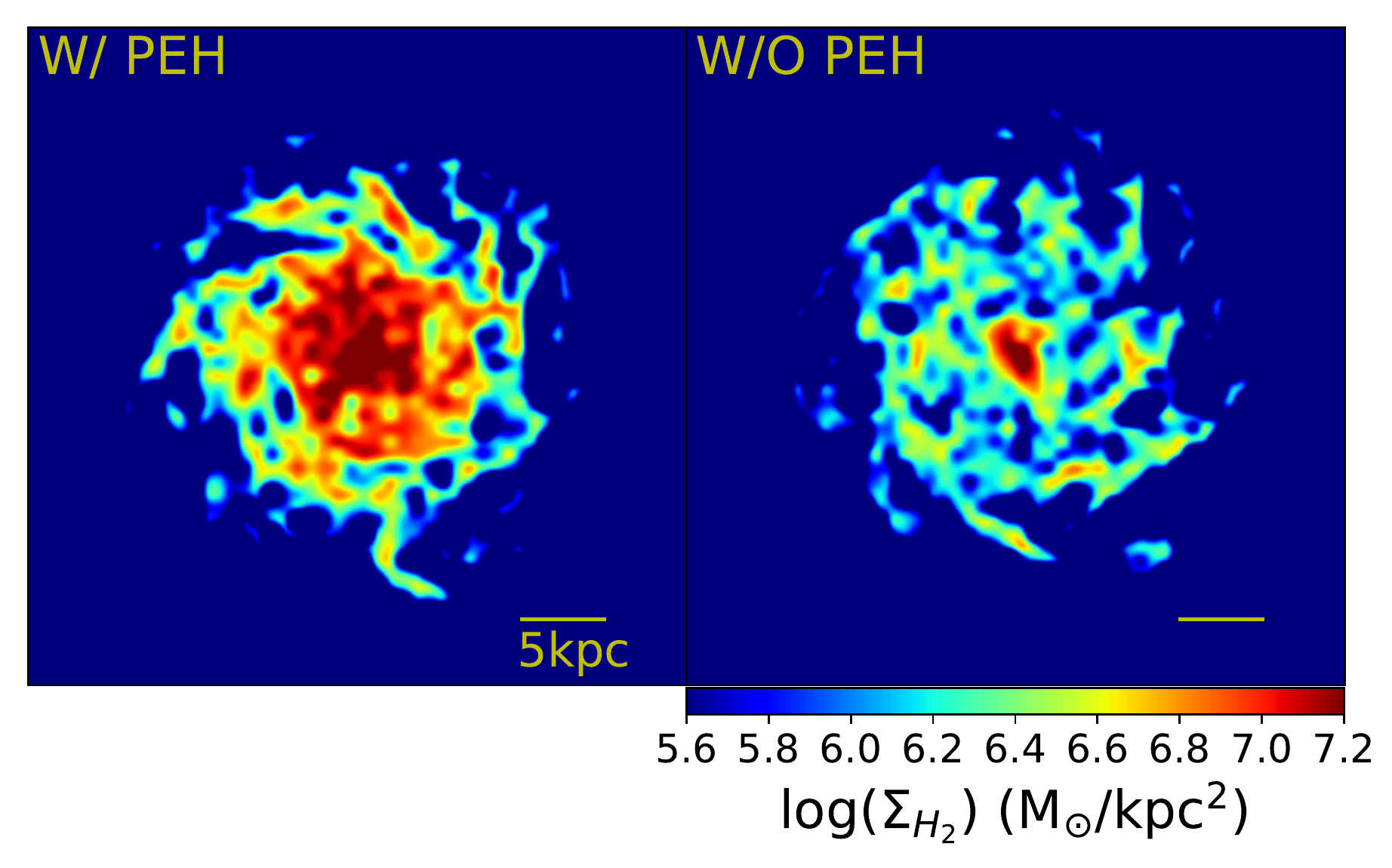}\par
    \includegraphics[height=5cm,width=8.5cm]{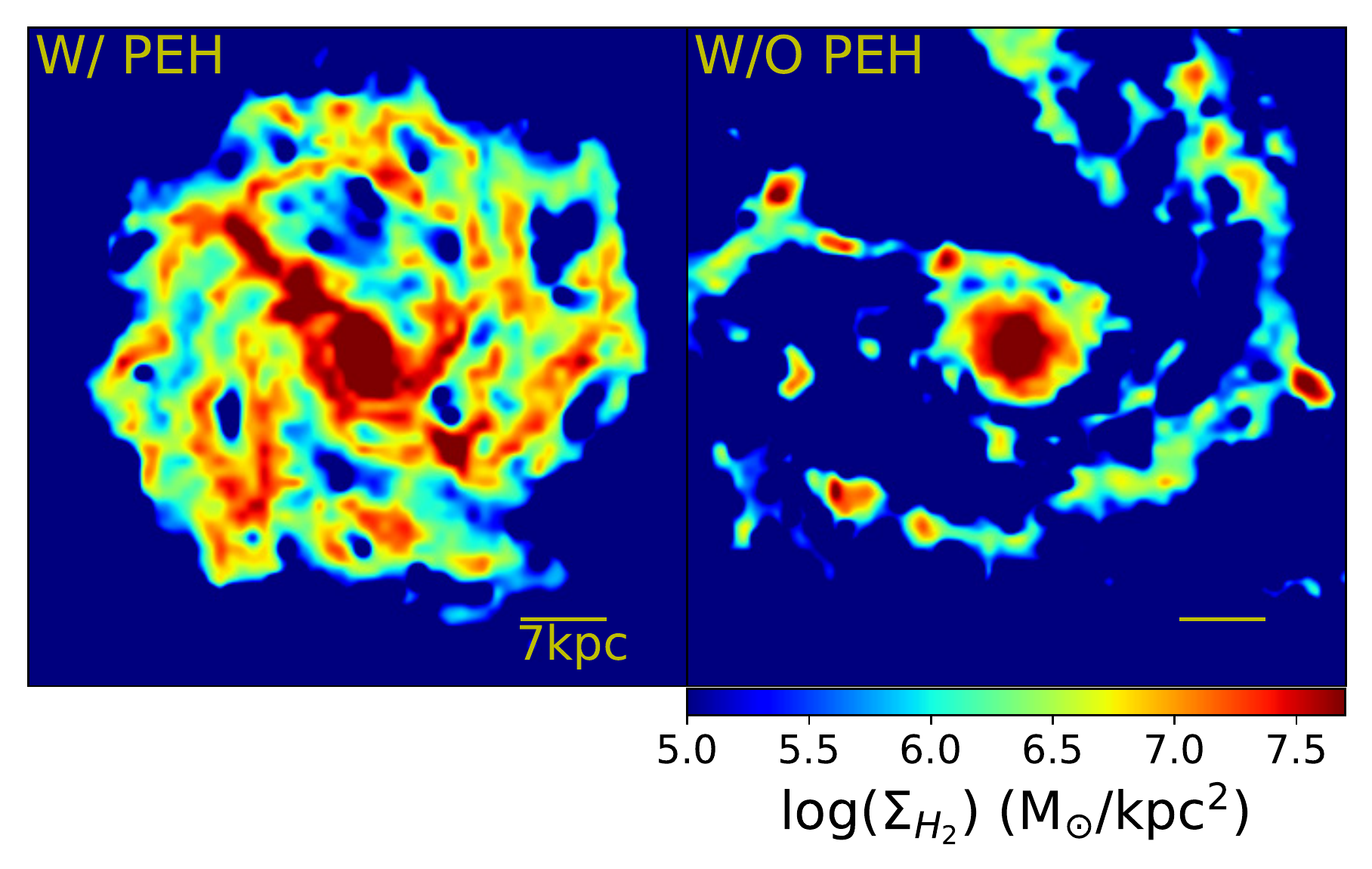}\par 
    \end{multicols}
    \begin{multicols}{2}
    \includegraphics[width=1.\linewidth]{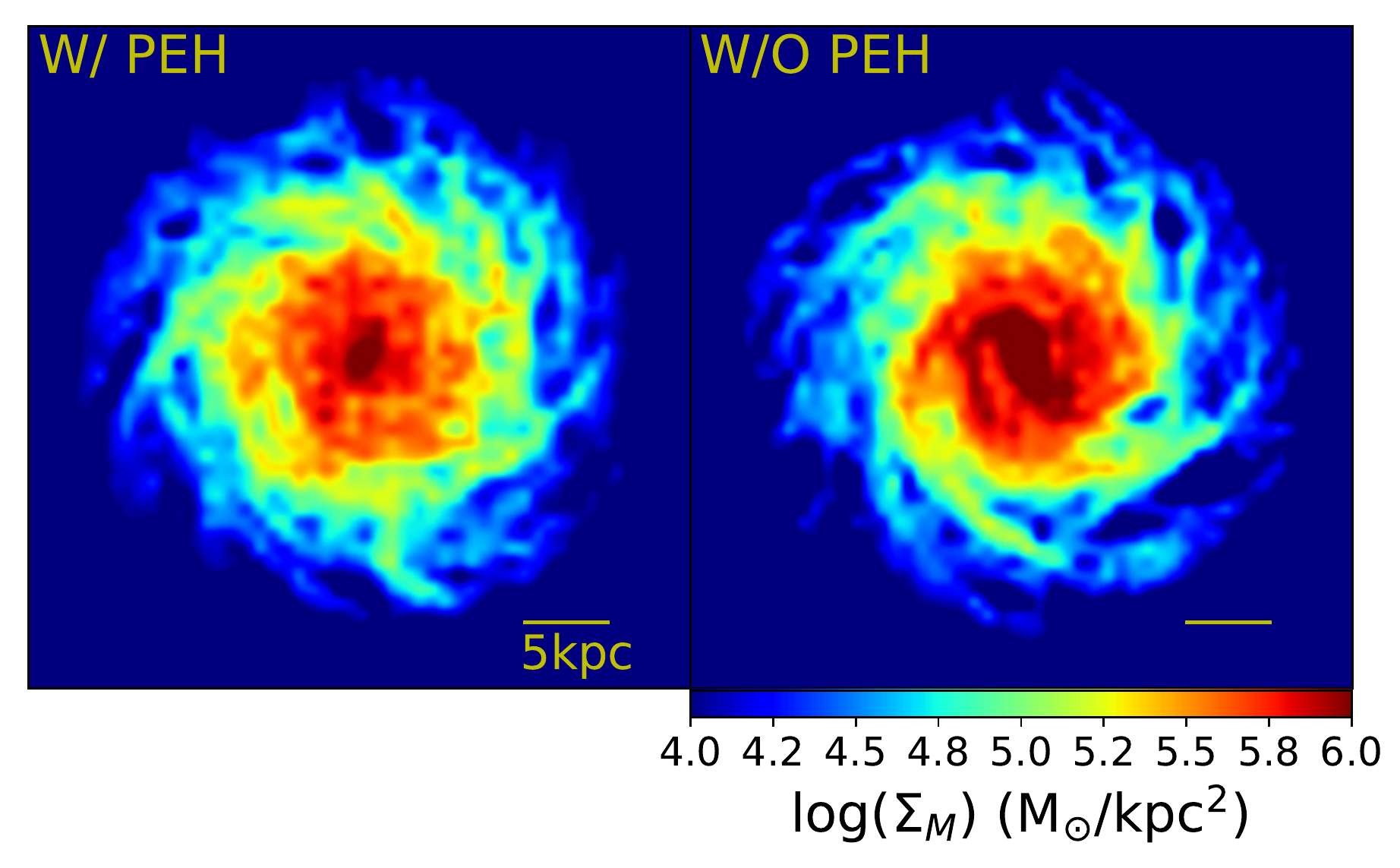}\par 
     \includegraphics[width=1.\linewidth]{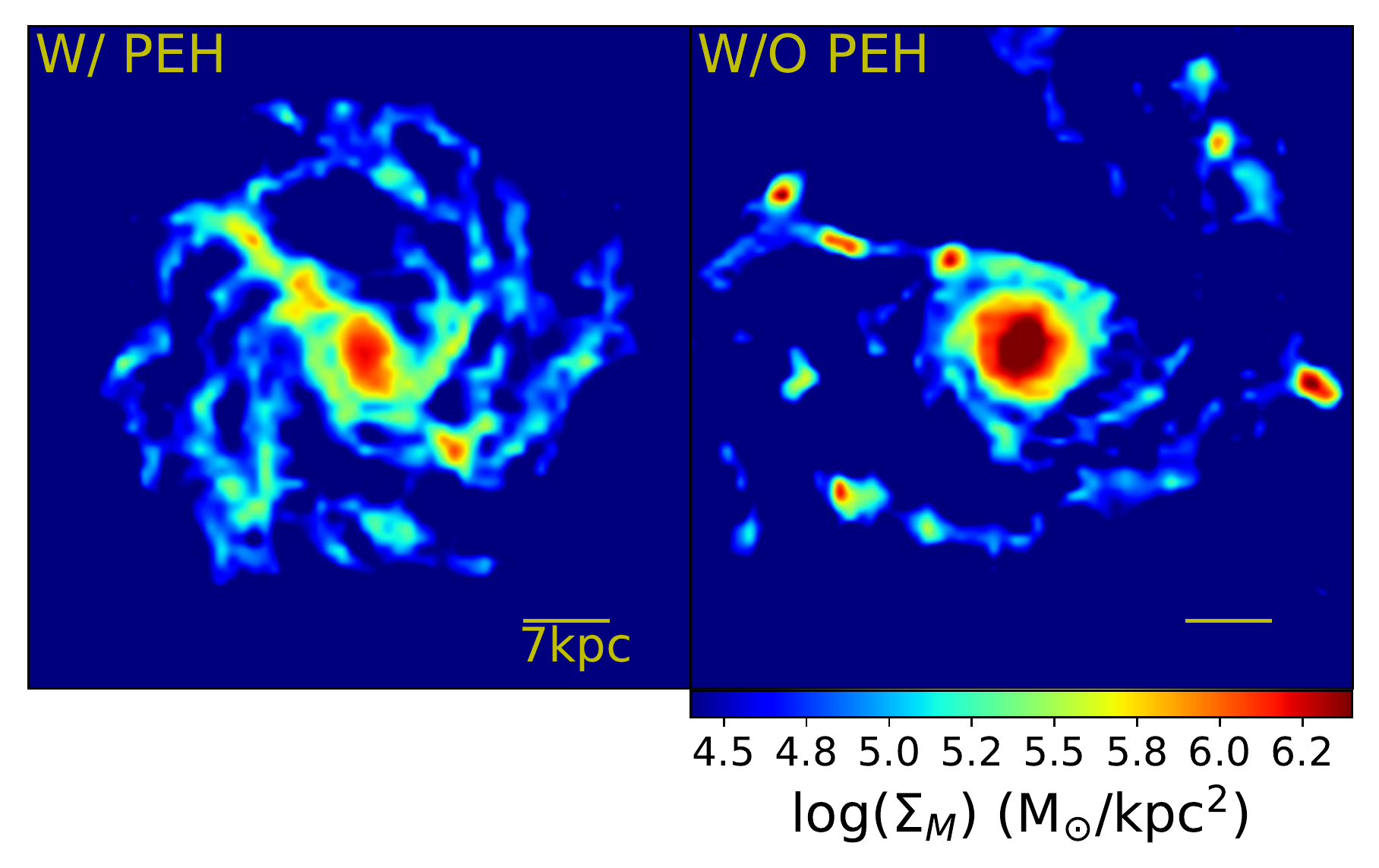}\par 
    \end{multicols}
    \begin{multicols}{2}
    \includegraphics[height=5cm,width=8.5cm]{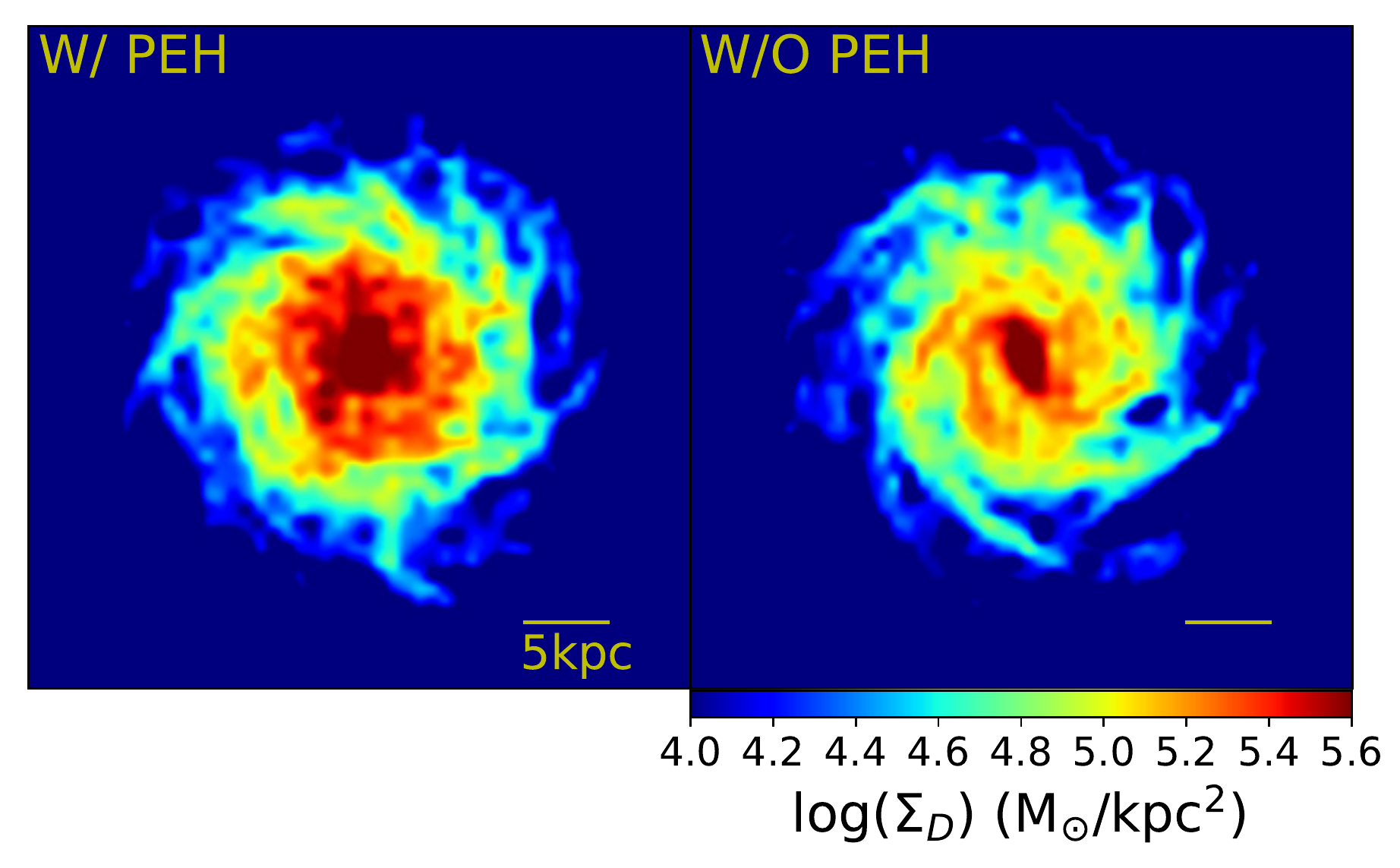}\par
    \includegraphics[height=5cm,width=8.5cm]{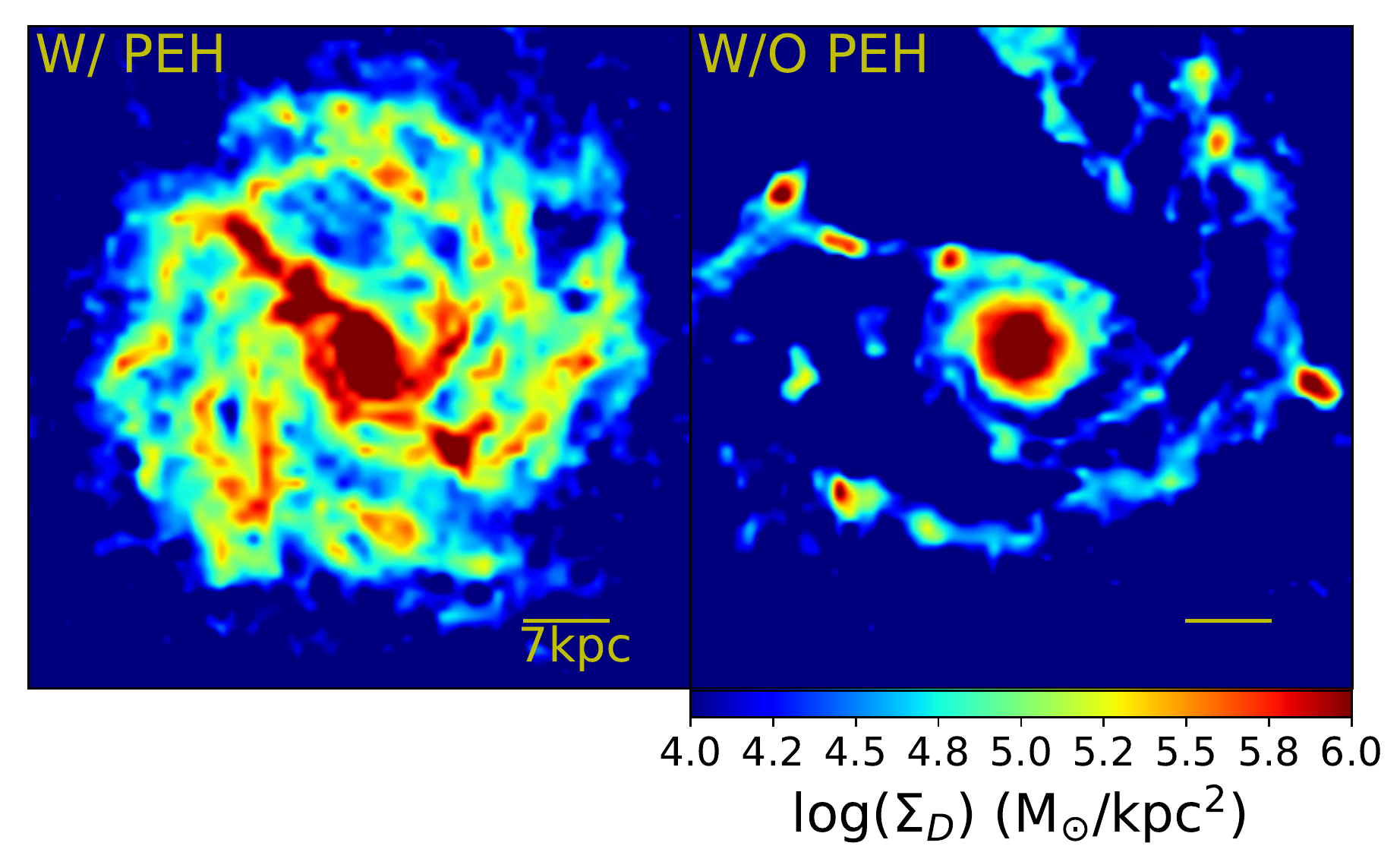}\par
    \end{multicols}

\vspace{-0.6cm}
\caption{The $xy$ projection of the mass surface density of the total gas (upper row), H$_2$ (middle--upper row), metals (middle--bottom row), and dust (bottom row) ($\Sigma_{G}$, $\Sigma_{H_2}$, $\Sigma_{M}$, and $\Sigma_{D}$, respectively) in logarithmic scale at T = 1 Gyr. The left column shows M1 (left subfigures in each panel) and M2 (right subfigures in each panel) models. The right column shows M3 (left subfigures in each panel) and M4 (right subfigures in each panel) models.}
\vspace{-0.3cm}
\label{fig4}
\end{figure*}

Fig. \ref{fig4} shows the $xy$ projections of the mass surface density of the total gas (upper row), H$_2$ (middle--upper row), metals (middle--bottom row), and dust (bottom row) ($\Sigma_{G}$, $\Sigma_{H_2}$, $\Sigma_{M}$, and $\Sigma_{D}$, respectively) in logarithmic scale at T = 1 Gyr. The left column shows models M1 (left subfigures in each panel) and M2 (right subfigures in each panel). The right column shows models M3 (left subfigures in each panel) and M4 (right subfigures in each panel).

The extra heat supplied to the ISM through PEH causes the 3D density of the individual gas particles ($n$) to drop which in turn suppresses star formation. Accordingly, M1 and  M3 models after 1 Gyr of evolution have 6\% and 31\% higher amount of gas in total compared to the gas in M2 and M4 models, respectively. The process also influences H$_2$ and dust abundances, as the relatively quiet environment (less star formation and SNe going off) allows H$_2$ to continue forming on dust grains (Cazaux \& Tielens 2004; Fukui \& Kawamura 2010) and dust grains to continue growing in molecular clouds (Savage \& Sembach 1996; Jones 2000; Sofia 2004). The result is 55\% and 43\% higher amount of H$_2$ and 33\% and 46\% higher amount of dust in M1 and M3 models, respectively. Metals are affected in such a way that models without PEH have a higher amount of metals. This is a result of the higher SFRs (more metals produced by stars) and the less/more efficient dust growth/destruction in these models compared to models with PEH. Thus, the gas, H$_2$, and dust surface densities are higher in models M1 and M3 compared to models M2 and M4, while M2 and M4 have higher metals surface densities. The increase/decrease in the total mass of the H$_2$, dust, and total gas, in 1 Gyr (the mass growth rate) is given in Table 4.

\begin{figure*}
\begin{multicols}{2}
       \includegraphics[height=6cm, width=8cm]{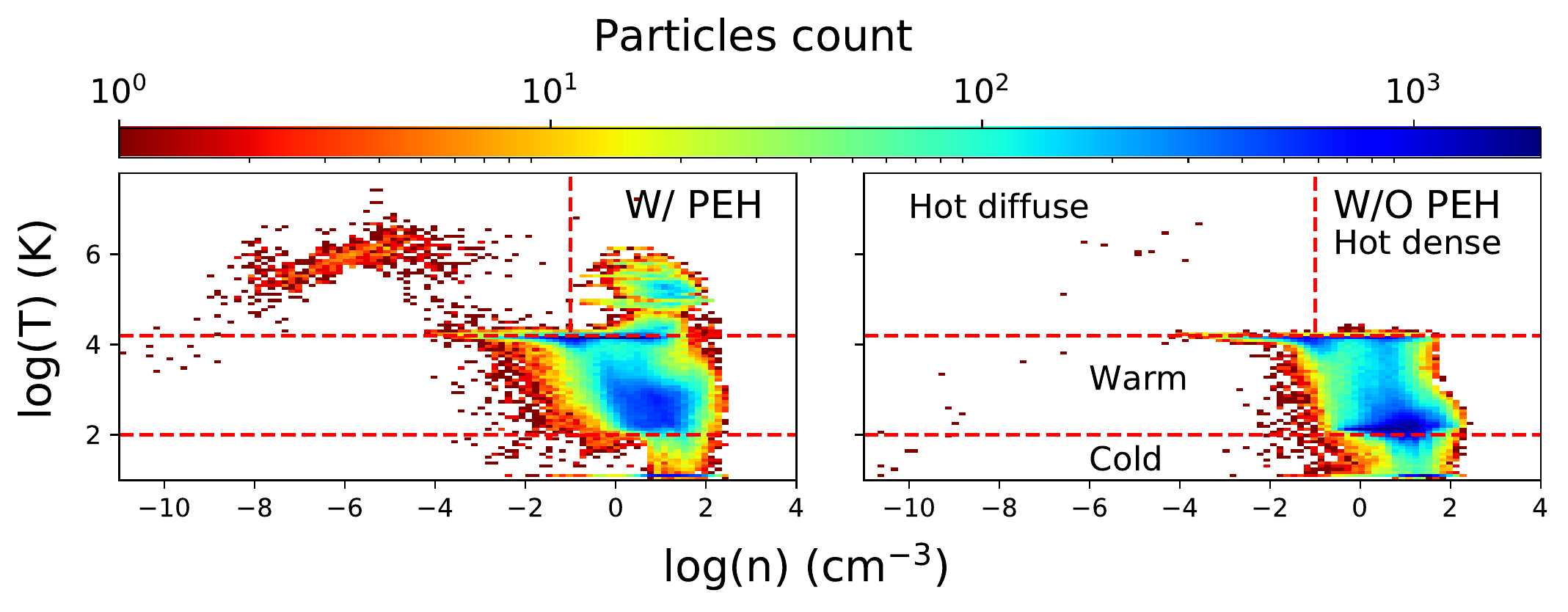}\par 
        \includegraphics[height=6.cm,width=8cm]{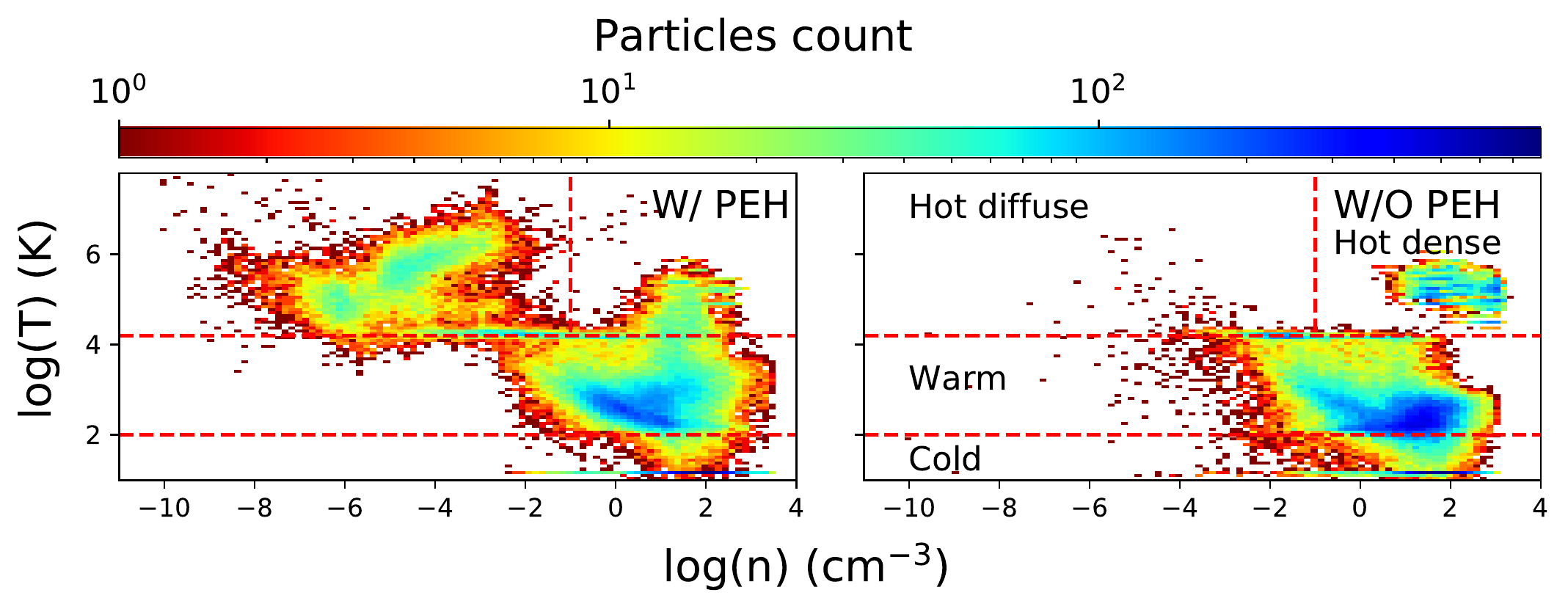}\par
        \end{multicols}
\vspace{-.5cm}
\caption{2D histograms of the temperature versus density at T = 1 Gyr. Left panel shows models M1 (left subfigure) and M2 (right subfigure). Right panel shows models M3 (left subfigure) and M4 (right subfigure). The vertical and horizontal dashed lines roughly illustrate regions of the phase diagram where the different ISM phases are contained according to their temperature.}
\label{fig5}
\end{figure*}

\begin{figure*}
\begin{multicols}{2}
    \includegraphics[height=6cm,width=8.cm]{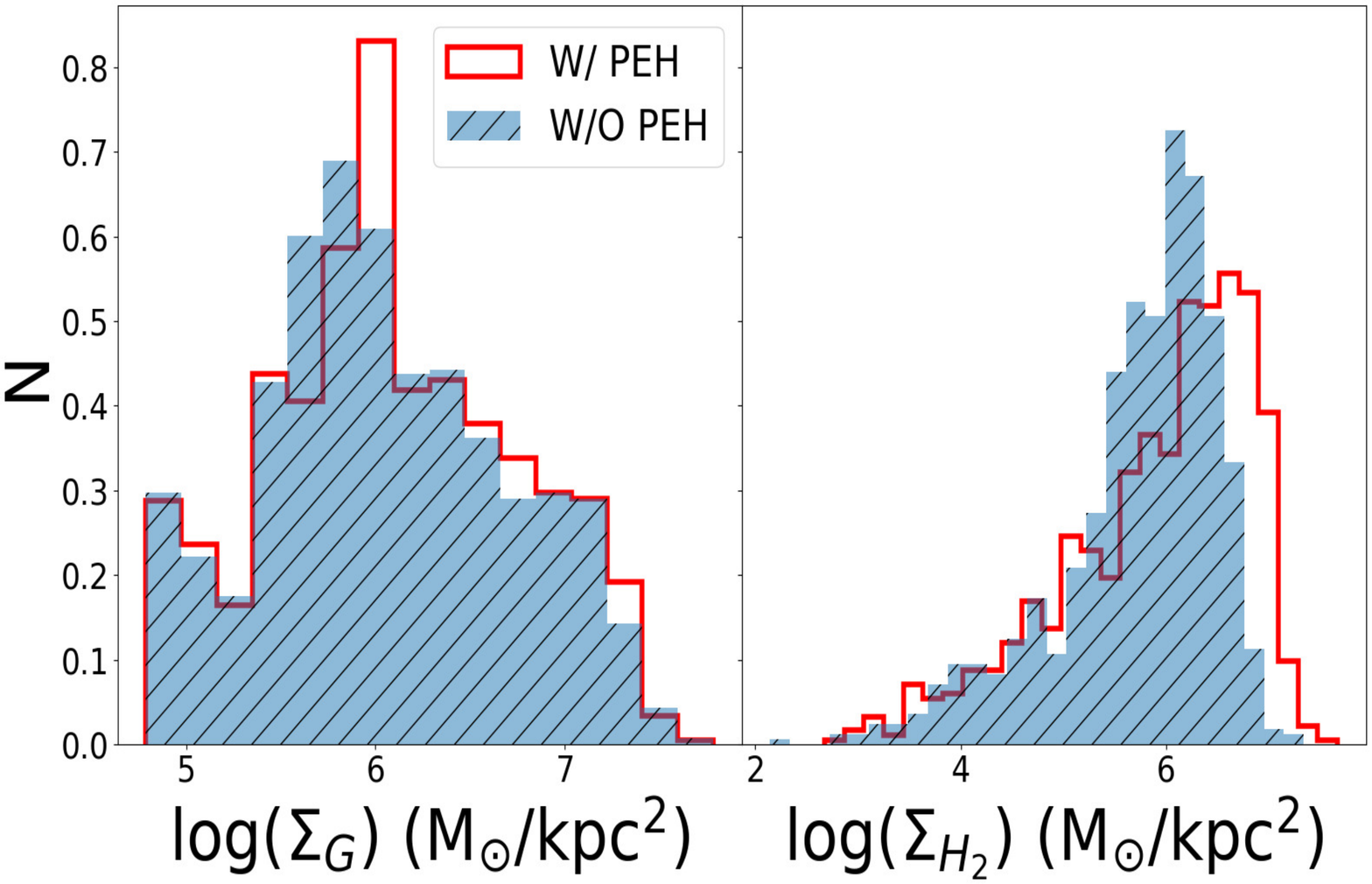}\par 
    \includegraphics[height=6cm,width=8.cm]{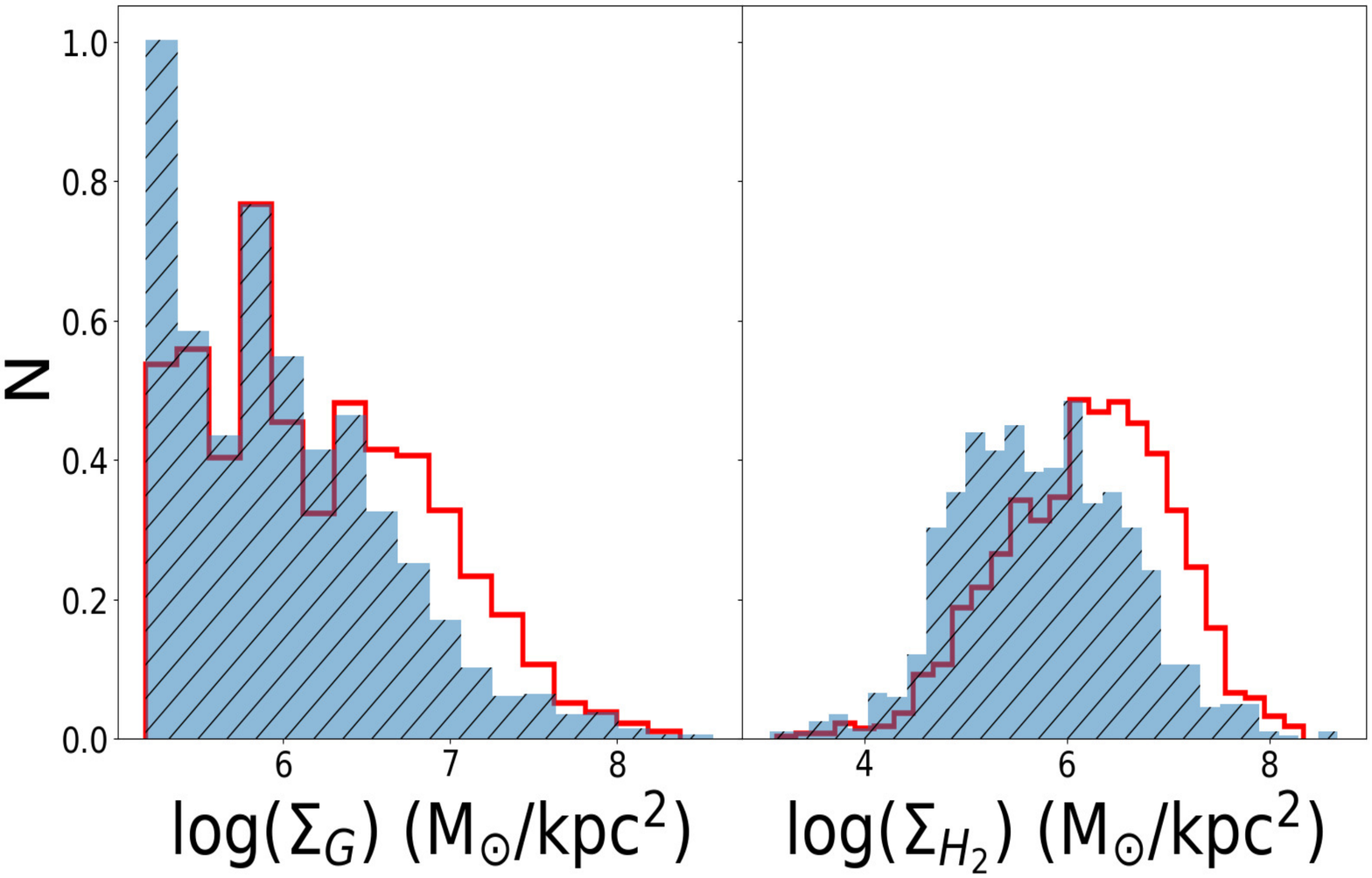}\par 
    \end{multicols}
       \vspace{-0.6cm}
\begin{multicols}{2}
    \includegraphics[height=6cm,width=8.cm]{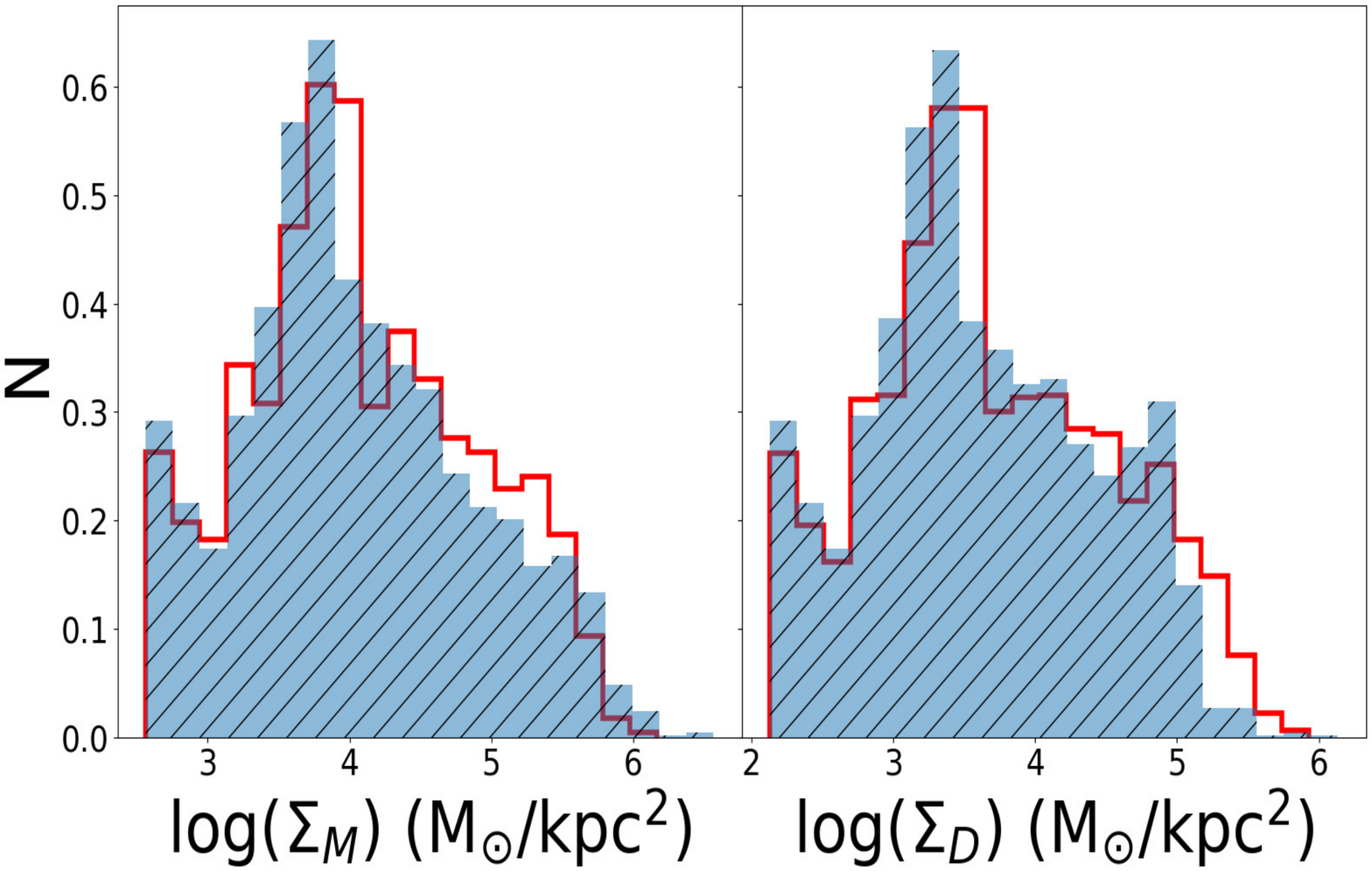}\par 
    \includegraphics[height=6cm,width=8.cm]{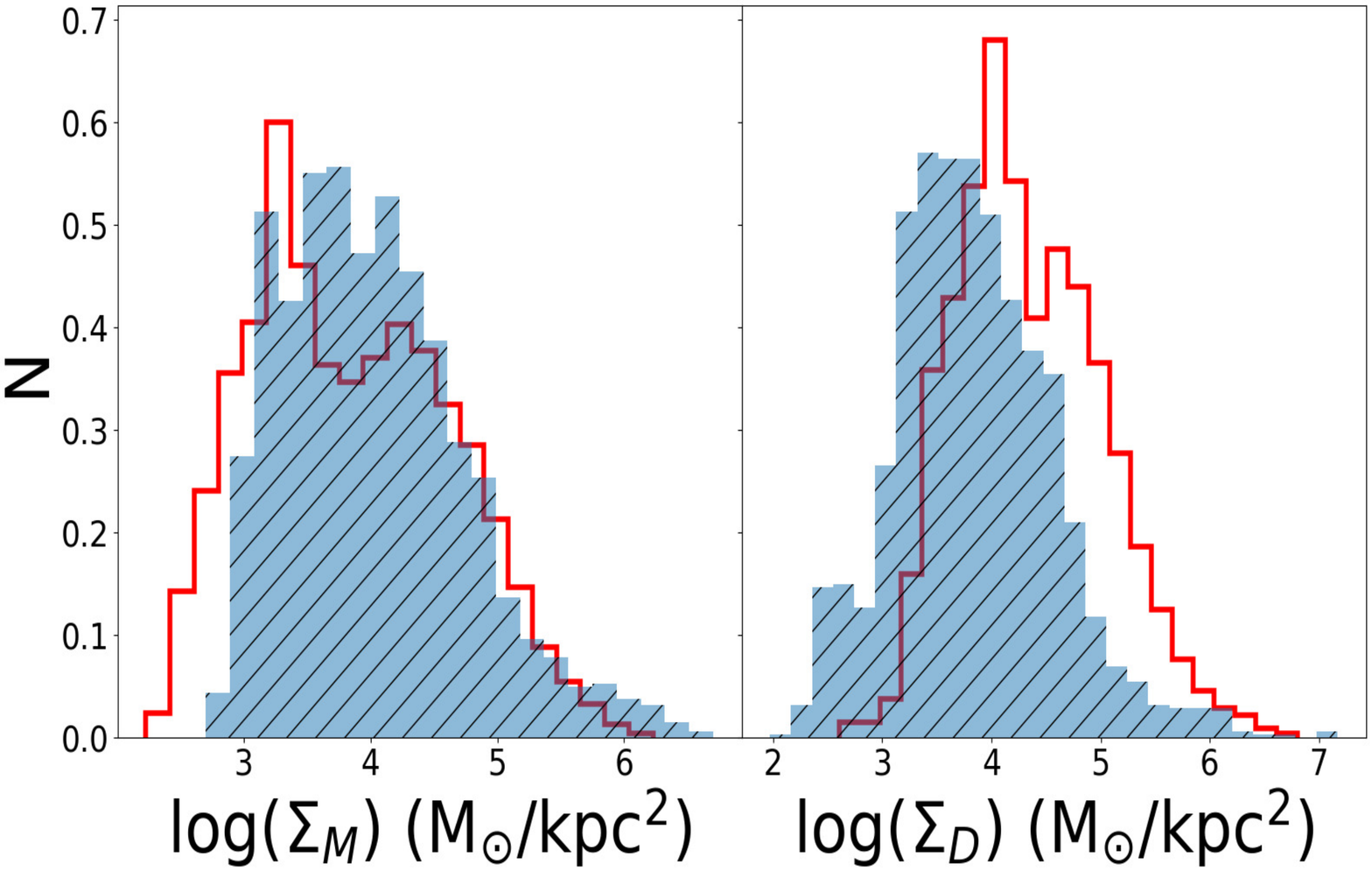}\par 
    \end{multicols}
       \vspace{-0.6cm}
\caption{1D normalized histograms of the total gas and H$_2$ surface densities (top row), metals and dust surface densities (bottom row) at T = 1 Gyr. Left column shows models M1 (red--empty histograms) and M2 (blue filled--hatched histograms). Right column shows models M3 (same as M1) and M4 (same as M2).}
\label{fig}
\end{figure*}

Furthermore, the $xy$ projection of the total gas, H$_2$, and dust in M1 model show multiple high--density cores apart from the central region, while metals projection does not show such cores since at the position of these cores metals are used up by dust growth. The area and density contrast between these cores and their background are probably smaller than the characteristic values of clumps. On the contrary, the $xy$ projections in M2 do not show as many high--density regions. Although gas distribution in M1 model shows a rather small bar or a boxy/peanut--shaped bulge (Berentzen et al. 2007) compared to the bar seen in M2, the stellar distributions show spiral arms in both models, a bar in M1, and a boxy--shaped bulge in M2. It is difficult to conclude that one galaxy/galaxy--model is clumpier than the other just by visual inspection. In case we adopt the simple definition that clumps are high--density contrasts embedded in a low--density background covering an area of at least one square kpc (consistent with clump size in high--redshift clumpy galaxies), we find that only the H$_2$ projection in M2 shows distinct clumps.

All the different components of the ISM (HI, H$_2$, metals, and dust) show clumpy structure in M3 and M4 models with M4 model showing more prominent clumps compared to their background. Young stellar populations associated with these gas clumps indicate the efficient star formation therein. M3 model also shows a disrupted bar with a bulge compared to the bulge seen in M4 model. The bulge in M4 could, possibly, be a remnant of a bar that formed earlier and was destroyed due to the accretion of clumpy gas (bar self--destruction, Pfenniger \& Norman 1990; Norman et al. 1996; Kormendy \& Kennicutt 2004; Kormendy 2013) which is indeed present. The bulge could also be formed by migration of the gas clumps to the central regions (e.g., Elmegreen et al. 2008) without passing through the bar formation stage.  It is worth noting that the disk in M4 is a disrupted disk which means the inclusion of PEH (with efficiency as low as 0.003) helps in stabilizing gas--rich (f$_g$ $\geq$ 0.5) Milky way--like disks. At high redshifts, clumpy marginally stable disk galaxies (Hitschfeld et al. 2009; Puech 2010; Cacciato et al. 2012; Fisher et al. 2014) are found to have, on average, gas fractions between 0.4 and 0.6 (Tacconi et al. 2008; Daddi et al. 2008; 2010). Hence, disks in M3 and M4 could represent clumpy disks at high redshifts.

\begin{figure*}
\begin{multicols}{2}
\includegraphics[height=5cm,width=8.5cm]{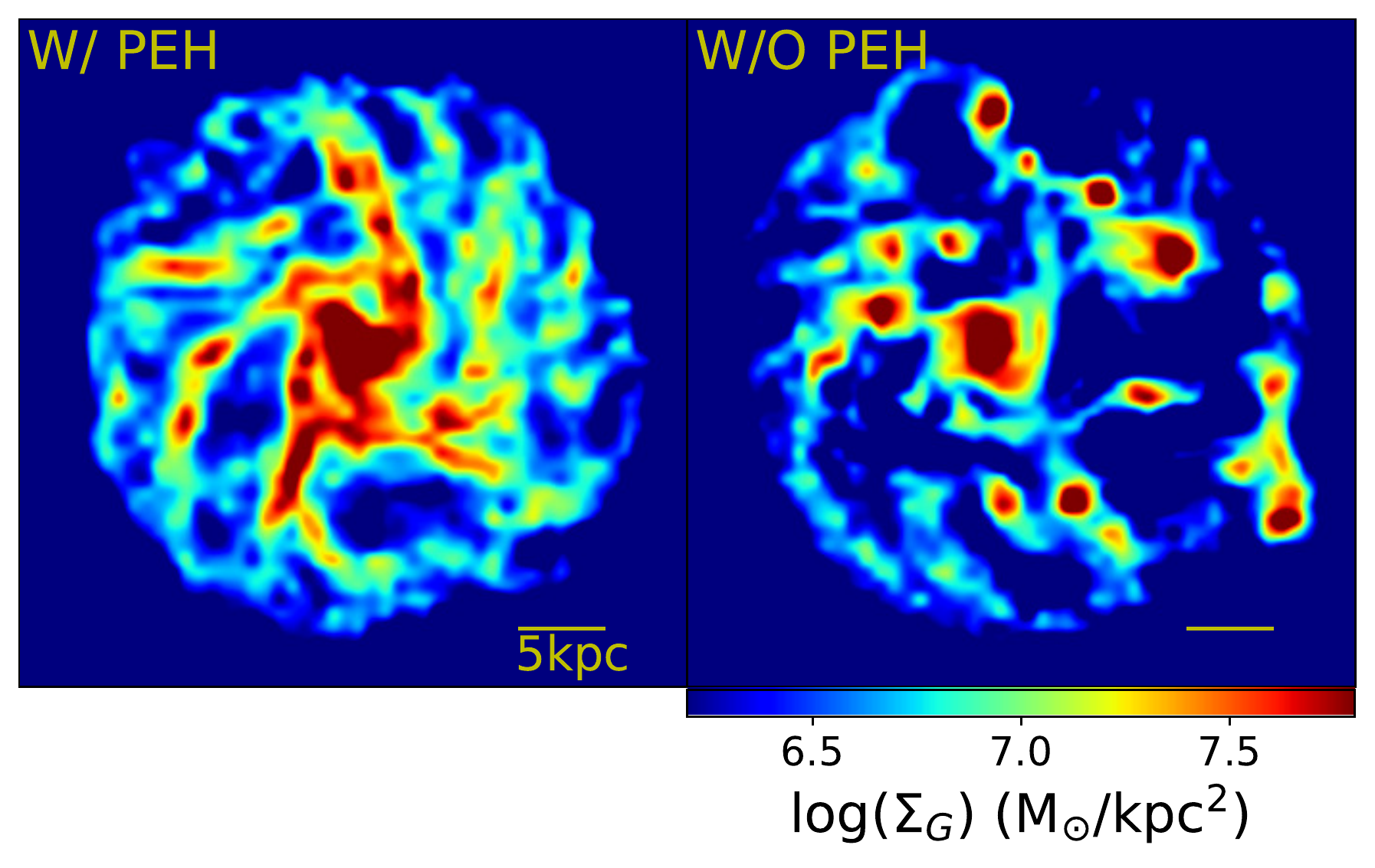}\par 
    \includegraphics[height=5cm,width=8.5cm]{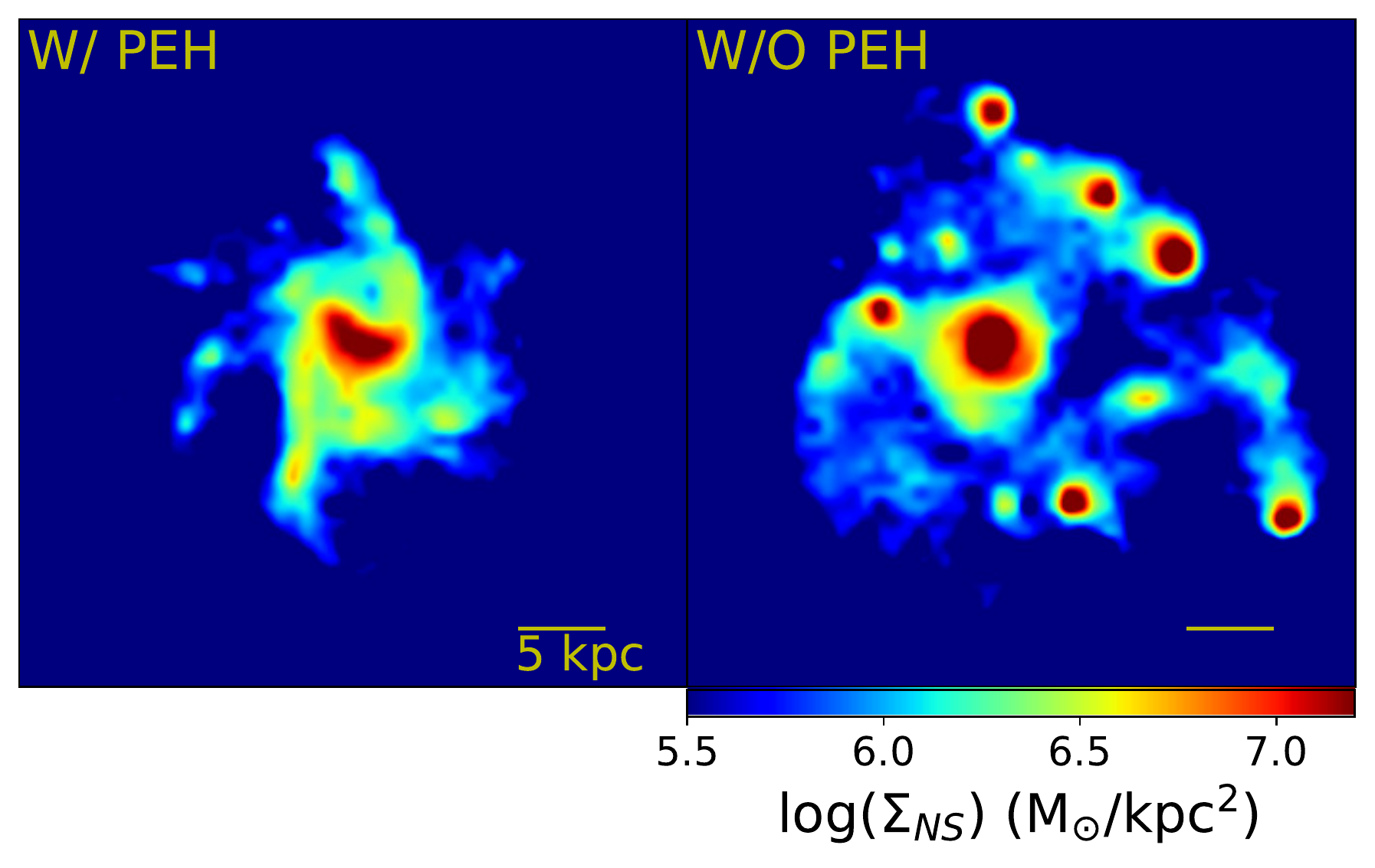}\par 
    \end{multicols}
\begin{multicols}{2}
	\includegraphics[height=5cm,width=8.5cm]{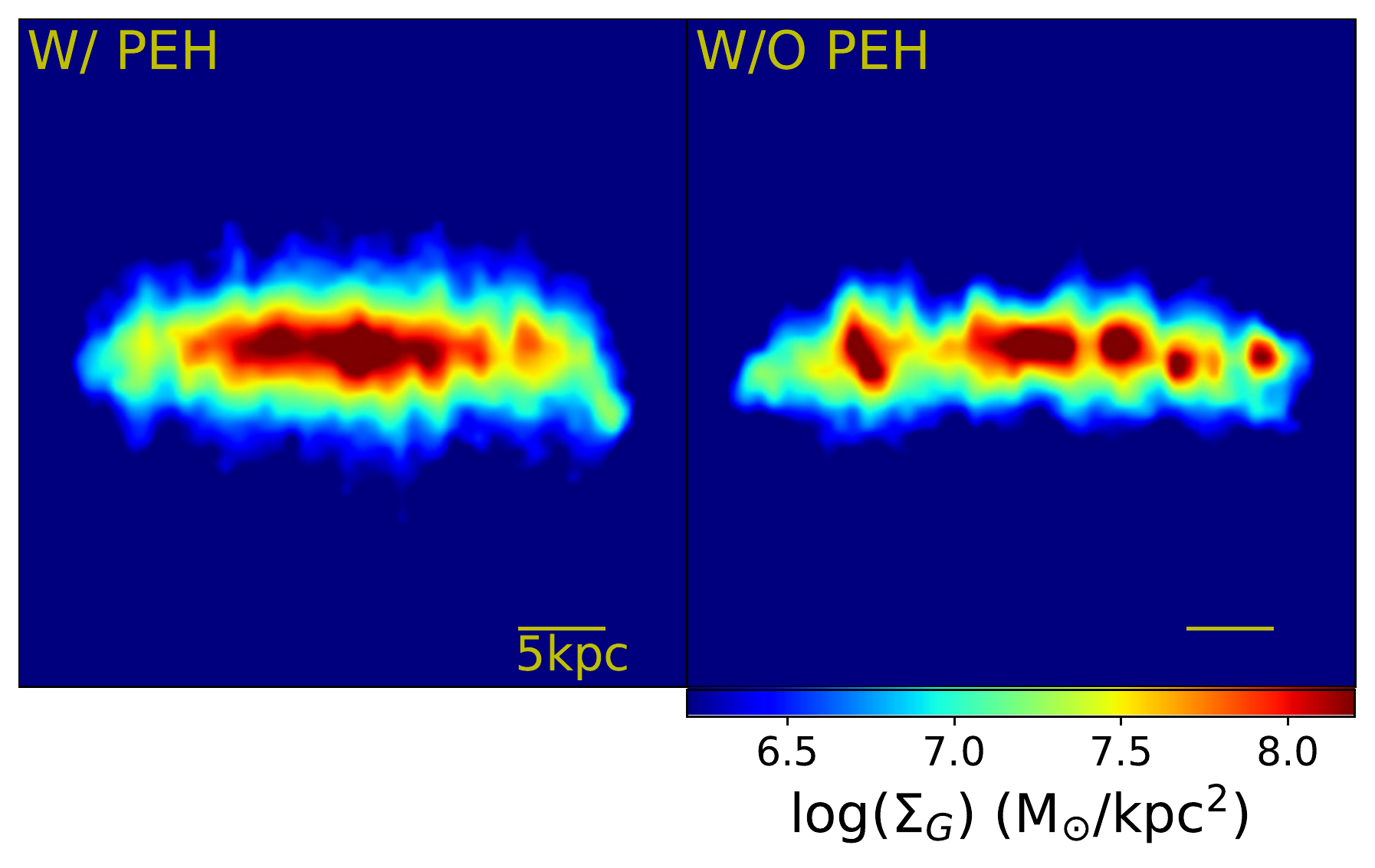}\par 
    \includegraphics[height=5cm,width=8.5cm]{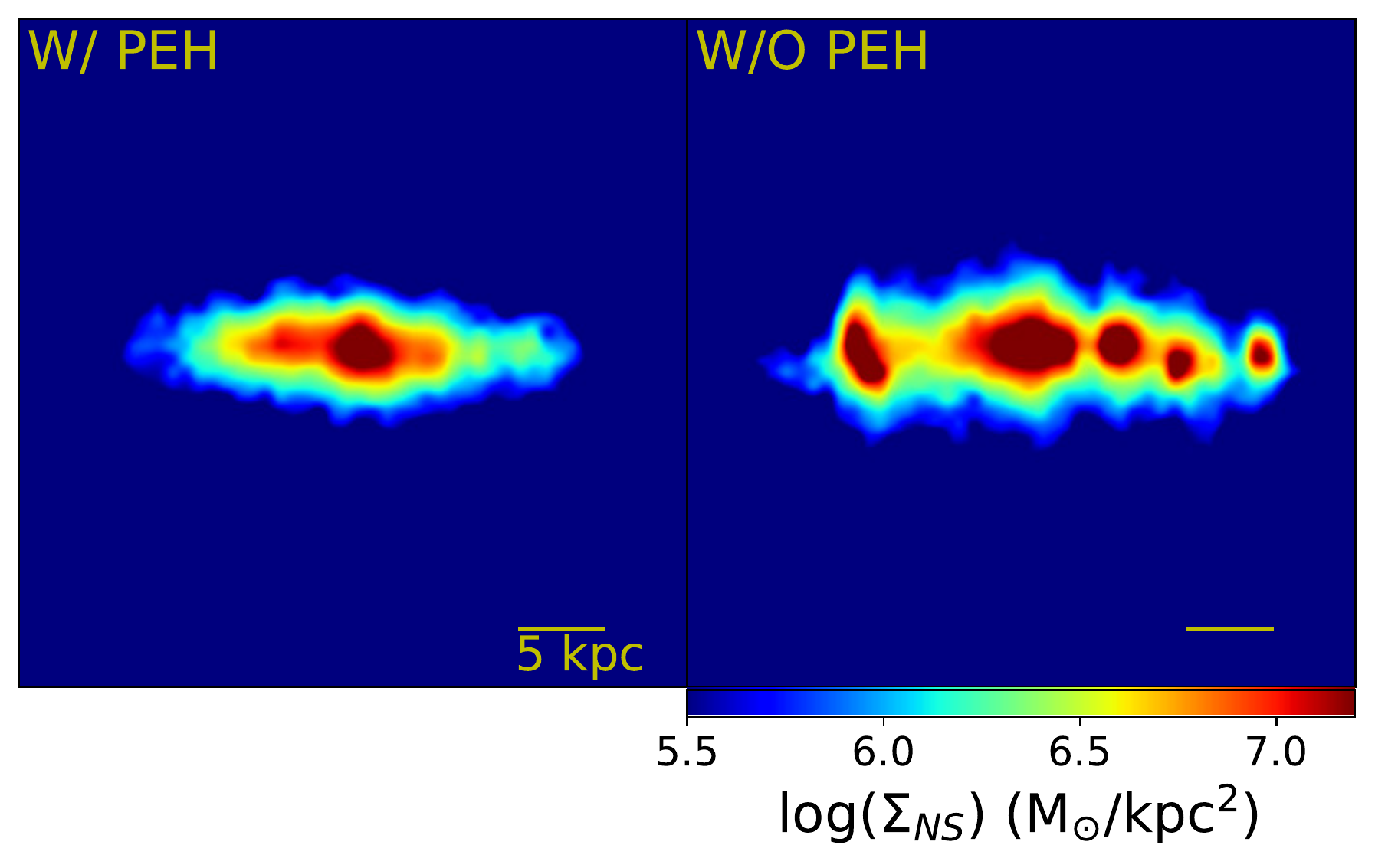}\par 
    \end{multicols}
\begin{multicols}{2}
    \includegraphics[height=6cm,width=8.5cm]{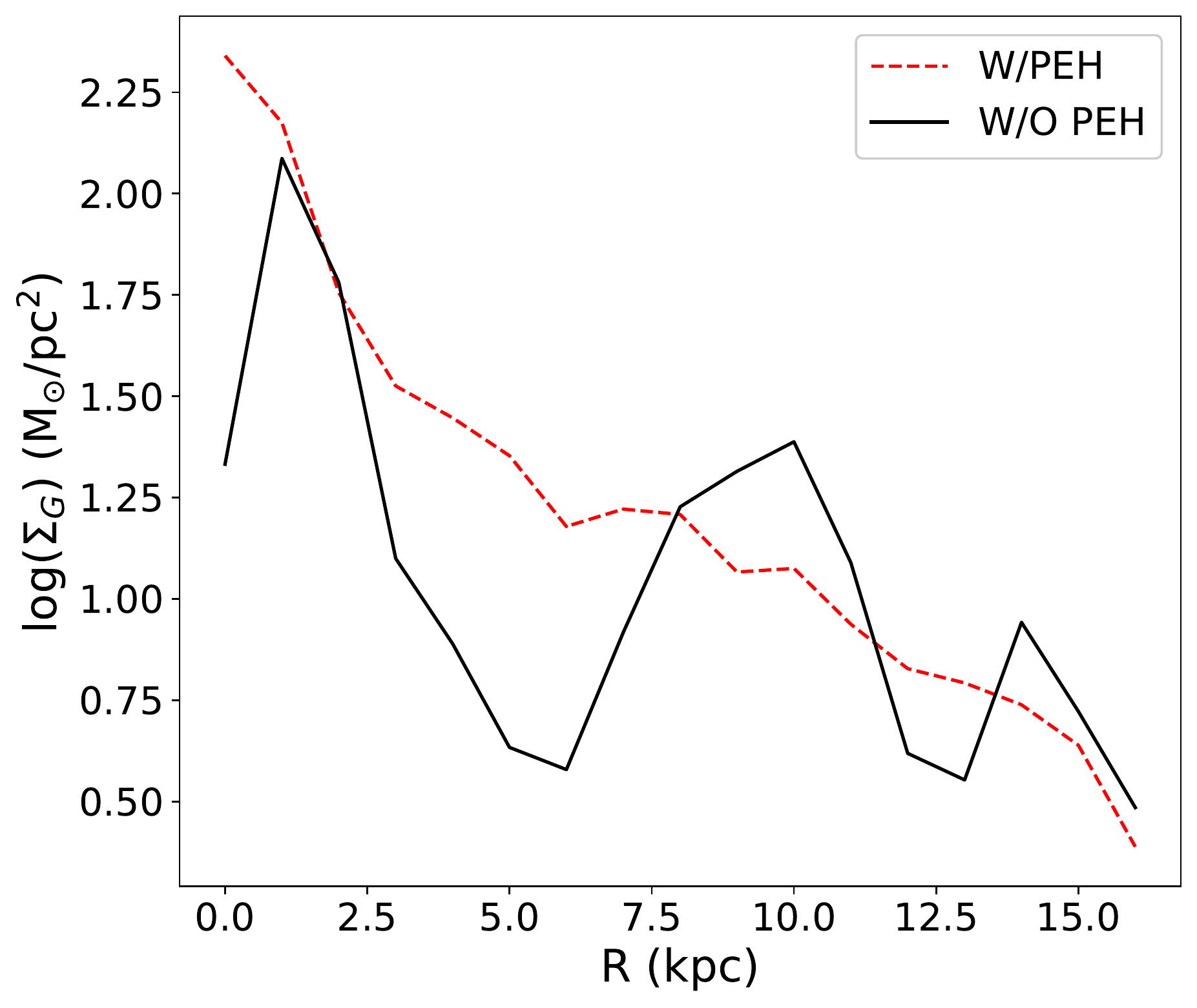}\par 
    \includegraphics[height=6cm,width=8.5cm]{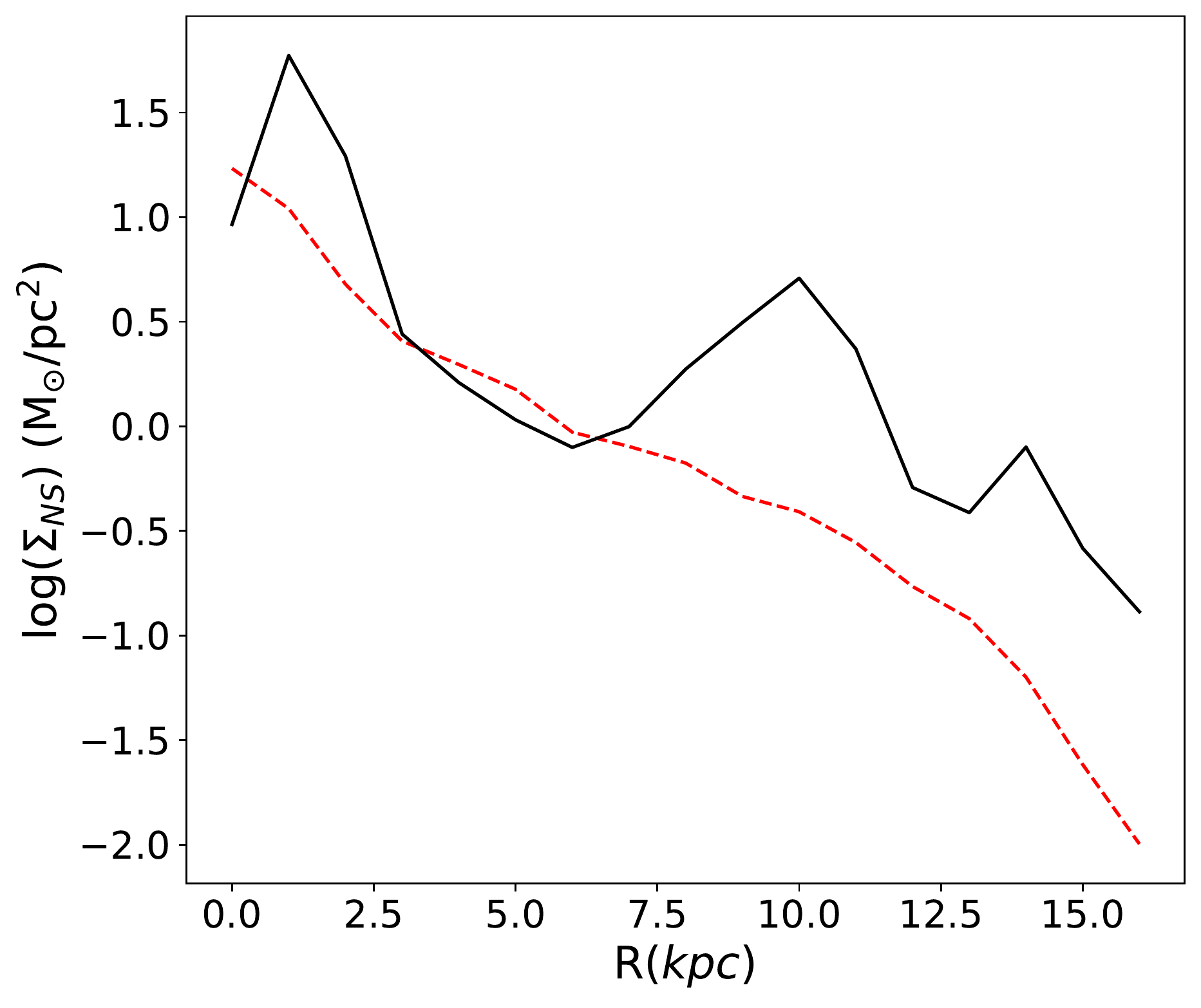}\par 
    \end{multicols}
    \vspace{-0.6cm}
\caption{$xy$ projection,  $xz$ projection, and the radial profiles of the total gas ($\Sigma_{G}$, left column from top to bottom, respectively) and the young stars ($\Sigma_{NS}$, right column from top to bottom, respectively). M19 is shown in the left subfigure of each panel in the top and middle rows and represented by the red dashed lines in the bottom row ($F_e$ = 0.05). M20 is shown in the right subfigure of each panel in the top and middle rows and represented by the black solid lines in the bottom row ($F_e$ = 0.0).}
%\vspace{-0.3cm}
\label{fig7}
\end{figure*}

\begin{figure}
\includegraphics[height=6.cm,width=9.cm]{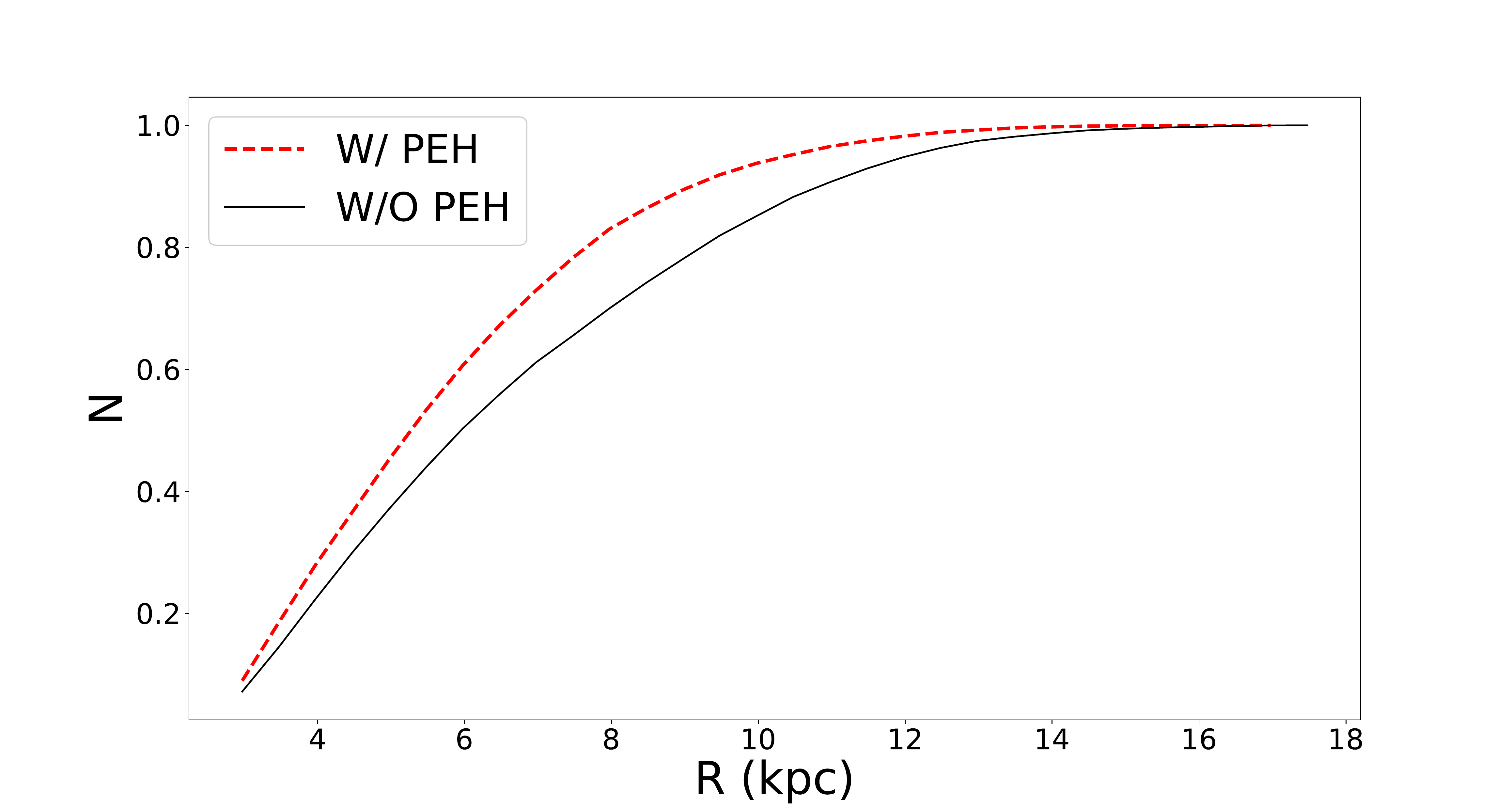}\par 
\caption{The cumulative histograms of the initial location on the disk (at T = 0) of the bulge gas particles (at T = 3 Gyr) excluding particles in the bulge at T = 0. The red dashed and black solid lines represent models M19 and M20, respectively.}
\vspace{-0.3cm}
\label{fig10}
\end{figure}

Effects of PEH on the ISM in models M1, M2, M3, and M4 are also depicted in Figs \ref{fig5} and \ref{fig}. Fig. \ref{fig5} illustrates the 2D histograms of the ISM temperature versus density at T = 1 Gyr. The left panel shows histograms of models M1 (left subfigure) and M2 (right subfigure). The right panel shows models M3 (left subfigure) and M4 (right subfigure). These phase diagrams show the multiphase nature of the ISM in the models where cold, warm, and hot gas phases co--exist. The vertical and horizontal dashed lines roughly illustrate regions of the phase diagram where the different phases are contained according to their temperature. The hot dense gas (T $>$ 10$^4$ K and n $>$ 1 $cm^{-3}$) seen in models M1, M3, and M4 is SNe shocked--heated gas located in the disk, while the low--density gas (n $<$ 10$^{-2}$ $cm^{-3}$) resides in the halo, mostly, ($|z| > 2$ kpc) and the think disk (1 $< |z| <$ 2 kpc). Models with PEH tend to have lower fractions of cold and warm gas, and a higher fraction of hot gas compared to the models without PEH, i.e. the ISM properties tend to shift to the upper right corner of the phase diagram when PEH is switched on. Moreover, the transition between the hot/warm gas with a temperature of around $10^4$ K and the warm/cold gas with a temperature of around $10^2$ K is smoother in the presence of PEH.

ISM in other simulations without PEH (e.g. Forbes et al. 2016; Hu et al. 2017) has hot gas fraction that is likely higher compared to our models without PEH. Moreover, in other studies with and without PEH there is no to little dense hot gas ($T > 10^4$ K and $n > 1$ $cm^{-3}$) formed in the models (e.g. Hu et al. 2016; Forbes et al. 2016; Emerick et al. 2019) apart from the model in Capelo et al. (2018) study. Capelo et al. (2018) noted that this gas is kept artificially hot because of the cooling scheme they implement. One interesting diagram is shown in Aoyama et al. (2017; Fig. 2), which is similar to our models with PEH although their model is without PEH.  It is also interesting to see the suppression of hot gas formation ($T > 10^4$ K) when PEH is included in Forbes et al. (2016) (however they have a cooling problem in their models; see Hu et al. 2017) models. In Forbes et al. (2016) and Tasker (2011) studies, PEH seems to suppress the scatter in the phase diagram.

Fig. \ref{fig} illustrates normalized 1D histograms of the total gas and H$_2$ surface densities (top row), metals and dust surface densities (bottom row) at T = 1 Gyr. The left column depicts histograms of models M1(red--empty histograms) and M2 (blue filled--hatched histograms). The right column depicts models M3 (same as M1) and M4 (same as M2). Histograms are normalized in such a way that the total area under the histogram is unity. The total gas and metals surface densities have similar distributions in models with and without PEH and f$_g$ = 0.1 (left column), while H$_2$ and dust surface densities distributions are slightly shifted towards higher values in M1 (with PEH). The differences in the ISM properties between models with and without PEH become more considerable in models with f$_g$ =  0.5 (right column). In this case, distributions of the total gas and metals surface densities are also different. Models with f$_g$ = 0.03 and  0.3 behave similar to the models in the left and right columns, respectively. Generally, the differences in the ISM properties between each pair of models (with and without PEH) diminish as the gas fraction decreases.

\subsection{PEH effects on the disk clumps}
Clumpy disk galaxies are observed not only in the high redshift universe but also in the local universe with less frequency though (Shibuya et al. 2016; Buck et al. 2017). Accordingly, we ran a pair of models that share the same characteristics with M3 and M4 but with baryon fraction 50\% less than what is present in M3 and M4 (i.e. f$_b$ = 0.03) to imitate present--day gas--rich galaxies, namely, M19 and M20 models. Fig. \ref{fig7} shows the $xy$ projection,  $xz$ projection, and the radial profiles of the total gas ($\Sigma_{G}$, left column from top to bottom, respectively) and the young stars ($\Sigma_{NS}$, right column from top to bottom, respectively). M19 is shown in the left subfigure of each panel in the top and middle rows and represented by the red dashed lines in the bottom row ($F_e$ = 0.05). M20 is shown in the right subfigure of each panel in the top and middle rows and represented by the black solid lines in the bottom row ($F_e$ = 0.0).

M19 shows smoother disk structure compared to the clumpy disk in M20 which is a result of violent disk instability. The massive gas clumps in M20 are associated with massive formation of young stars (right column, a common feature of clumps, Genzel et al. 2008; Forster Schreiber et al. 2009), unlike M19 where stellar clumps are less pronounced. While M20 shows clumps that do not follow a specific pattern and a slightly asymmetric disk (a feature of clumpy galaxies, Conselice et al. 2004), M19 shows a symmetric disk with clumps that are fragments of spiral arms (Inoue \& Yoshida 2018; 2019). Viewed edge--on, M19 disk resembles present--day galaxies with a triangular bulge, and M20 resembles chain galaxies (Cowie, Hu, \& Songaila 1995). The radial profiles show that M19 returns more gas in most of the disk compared to M20 since it forms stars at a lower rate. The clumpy structure is evident in both the gas and young stars profiles.

Migration of disk clumps to the central regions provides a viable mechanism for bulge formation (Noguchi 1999; Immeli et al. 2004; Dekel et al. 2009). After evolving M19 and M20 for a period of 3 Gyrs, M20 formed a bulge one order of magnitude more massive than the bulge in M19. Fig. \ref{fig10} shows the cumulative histograms of the initial location on the disk (at T = 0) of the bulge gas particles (at T = 3 Gyr) excluding particles in the bulge at T = 0, i.e. where the bulge particles located in the disk initially. Red dashed and black solid lines represent models M19 and M20, respectively. In the case of M20, gas particles from farther parts of the disk contribute to the bulge formation, unlike gas particles in model M19. The fraction of the disk particles that end up in the bulge at 3 Gyr is also higher in the case of M20.

\subsection{Gas fraction in the halo}
\begin{figure}
\vspace{0.5cm}
\includegraphics[height=5.5cm,width=8.5cm]{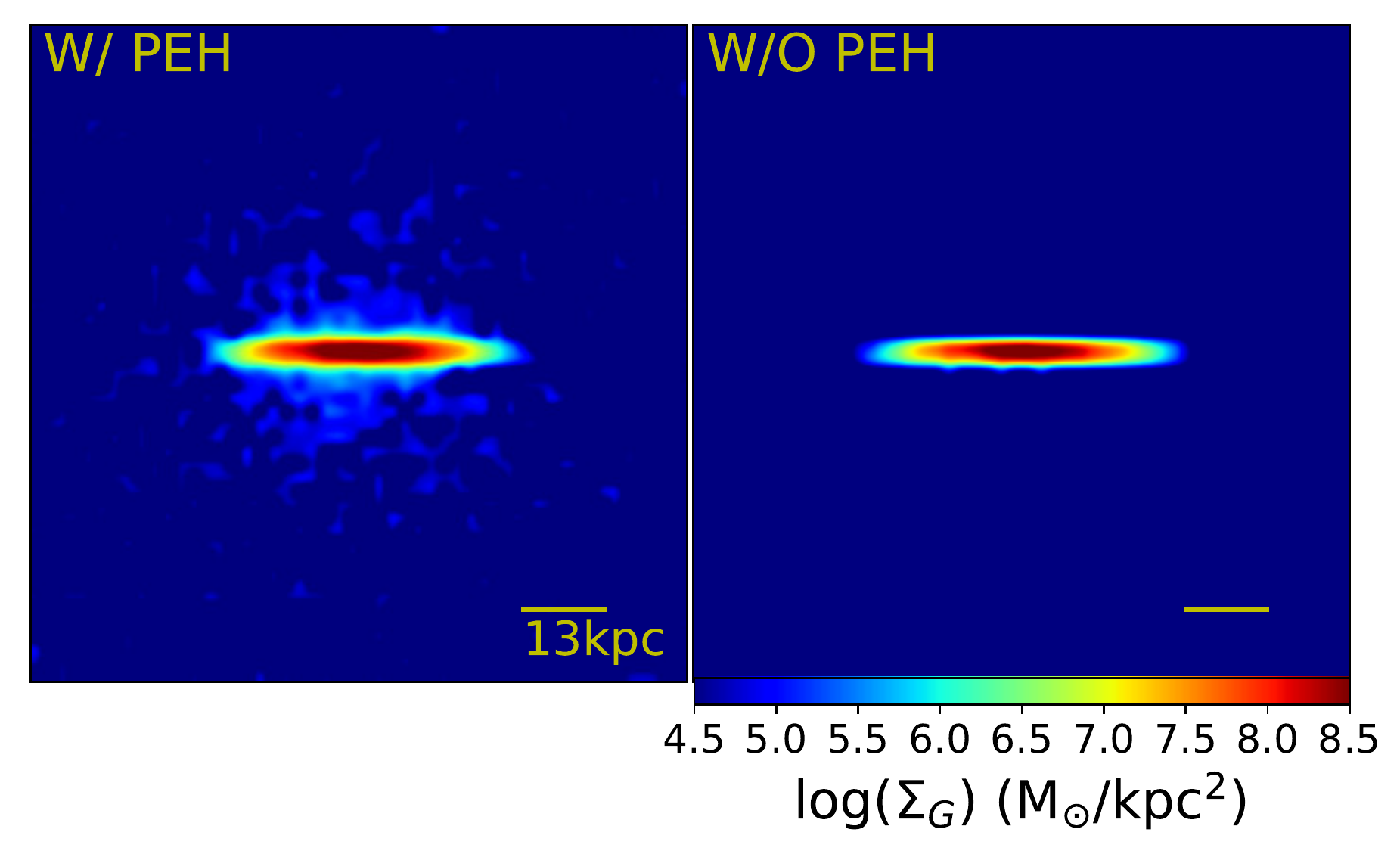}\par 
\caption{$xz$ projection of the gas surface density at T = 1 Gyr for models M1 (left) and M2 (right).}
\label{fig8}
\end{figure}

\begin{figure*}
\begin{multicols}{2}
    \includegraphics[height=6cm,width=8.cm]{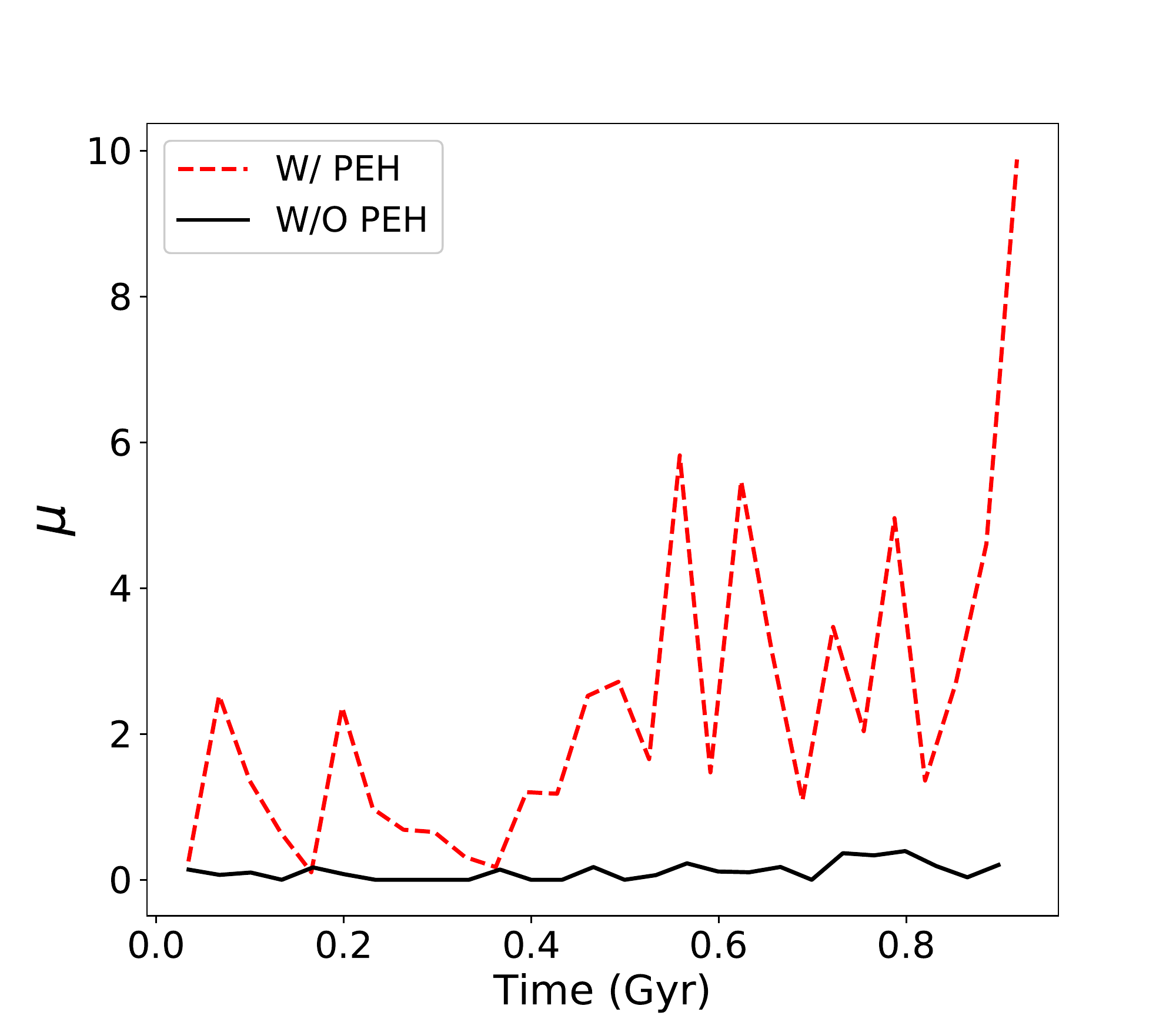}\par 
	\includegraphics[height=6cm,width=8.cm]{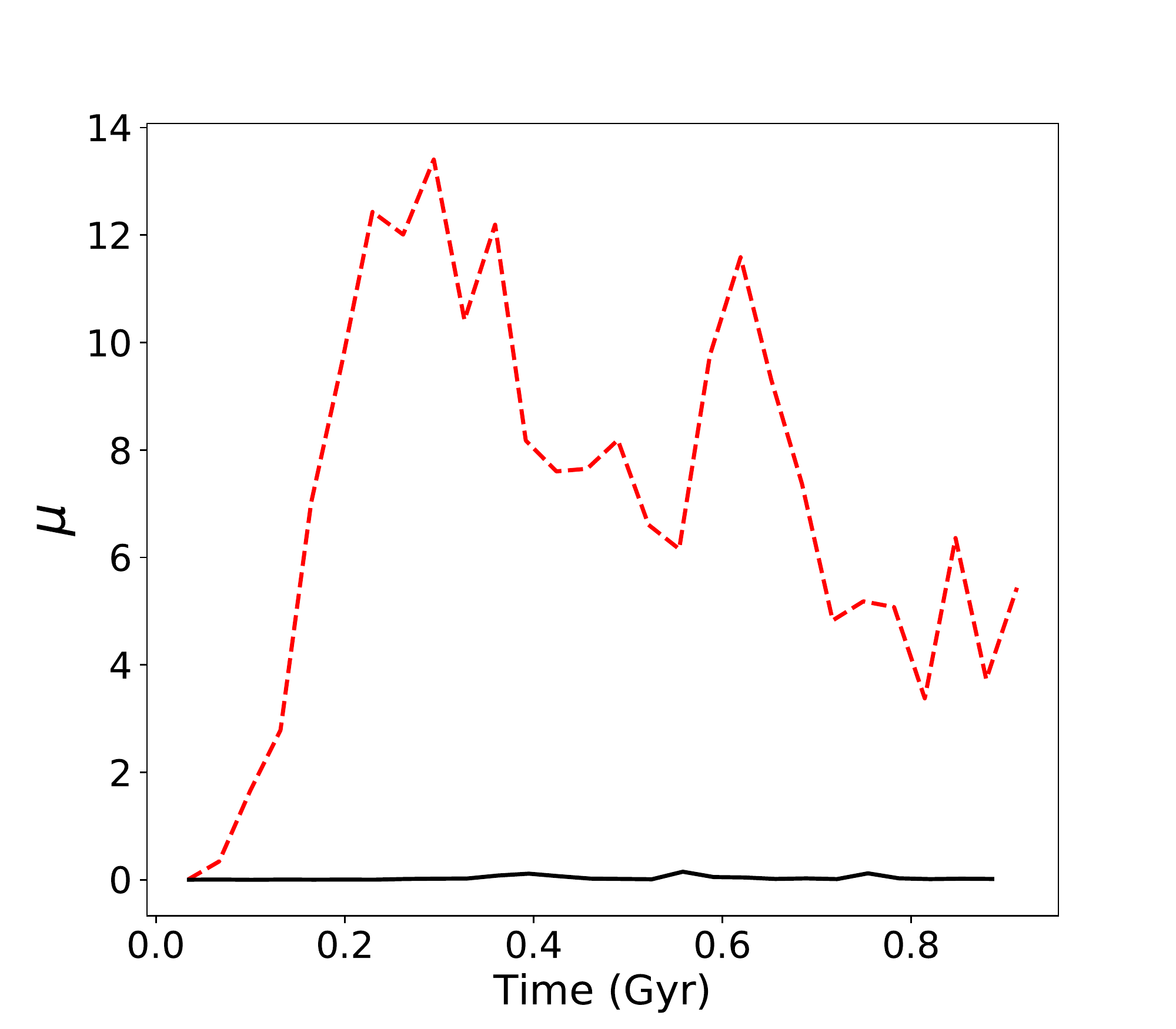}\par 
    \end{multicols}
    \vspace{-0.6cm}
\caption{Gas loading factor (ratio of the outflow rate to the SFR) as a function of time for models M1 and M2 (left panel), and M3 and M4 (right panel). The red dashed lines represent models M1 and M3, while the black solid lines represent models M2 and M4.}
\vspace{-0.3cm}
\label{fig8new}
\end{figure*}

Gas driven out to the halo by star formation activity and its subsequent SNe events (White \& Rees 1978; Dekel  \&  Silk  1986;  White  \&  Frenk 1991; see Veilleux et al. 2005; Naab \& Ostriker 2017 for overviews) is shown in Fig. \ref{fig8}. Fig. \ref{fig8} shows the $xz$ projection of the gas surface density at T = 1 Gyr for models M1 (left) and M2 (right). The disk where PEH in switched on (left) is slightly thicker and shorter compared to the disk where PEH is switched off, however, what is interesting is the amount of gas above the disk in each model. The higher amount of gas in the halo of M1 indicates a higher gas loading factor; accordingly, the time evolution of the gas loading factor ($\mu$) is investigated and is shown in Fig. \ref{fig8new} for models M1 and M2 (left panel), and M3 and M4 (right panel). Red dashed and black solid lines represent models M1 and M3, and M2 and M4, respectively. $\mu$ is the ratio between the gas outflow rate and SFR. The outflow rate ($\dot{M}_{out}$) is defined as the mass flux crossing a surface located at $|z|$ = 2 kpc per unit time and is calculated using the sum: $\dot{M}_{out} = \sum_{i} m_{g,i} \overrightarrow{v_{i}}.\hat{n}/\Delta x$ where $\hat{n}$ is the normal direction to the disk, $m_{g,i}$ and $\overrightarrow{v_{i}}$ are the mass and velocity of the i$th$ gas particle. Only particles with $\overrightarrow{v_{i}}.\hat{n} > 0$ are included in the sum, $\Delta x$ is taken to be 0.5 kpc. M1 and M3 have persistently higher gas outflow rates over time compared with M2 and M4, and $\mu$ shows that these outflow rates are quite significant compared to the SFR and could influence it.  Although the lower star formation activity in M1 and M3 compared with M2 and M4 implies that M2 and M4 have higher SNe rate which drives gas outflows, the ISM in M1 and M3 is hotter and less dense compared to the ISM in M2 and M4. For SNe to drive gas outflows efficiently, the ambient ISM needs to have low density (SNe efficiency decreases with increasing ambient density, see equation 10 in Naab \& Ostriker 2017; Hu et al. 2017; Hu 2019). Accordingly, we argue that PEH could influence the efficiency of SNe feedback since it influences the density of the ISM. Additionally, the gas driven out to the halo takes longer to cool down and fall back to the disk. Furthermore, the efficiency of SNe events in farther, less enriched, regions of the disk is also altered, resulting in less enriched gas outflows. 
%%%%%% TABLE1'
\begin{table}
\centering
\caption{A subset of the models presented in Table 3 where $\Delta M_{g}$, $\Delta M_{H_ 2}$, and $\Delta M_{D}$ are the mass growth rate ($\frac{M_{final} - M_{initial}}{M_{initial}}$) of the total gas, H$_2$, and dust, respectively, in 1 Gyr. $R_m$ is the change in the gas amount in the halo ($|z| > 2$ kpc) in 1 Gyr.}

\begin{tabular}{lllll}
\hline
{Model ID}  
 & $\Delta M_g$
 & $\Delta M_{H_2}$
 & $\Delta M_{D}$
 & $R_m$ ($\times 10^7 M_{\sun}$)\\
\hline
M1 &    $-0.070$& 31.8 & 0.34 & 1.40 $\pm$ 0.20\\
M2 &   $-0.130$& 13.8& $-0.12$&0.02 $\pm$ 1.80\\
M3 &   $-0.065$& 47.0& 1.80& 135.00 $\pm$ 22.00\\
M4 &   $-0.350$ & 26.4& 0.49& 7.30 $\pm$ 5.20\\
M5 &   $-0.072$& 54.5& 1.90& 34.00 $\pm$ 5.00\\
M6 &   $-0.240$& 37.0& 1.90&0.57 $\pm$ 2.90\\
M7 &  $-0.037$& 31.9& 0.25& 0.014 $\pm$ 0.50\\
M8 &   $-0.045$ & 22.7&0.16& 0.015 $\pm$ 0.50\\
M9 &    $-0.086$  & 38.6  & 0.60 & 0.43 $\pm$ 1.62\\
M10 &  $-0.112$  & 20.7  & 0.0006& 0.05 $\pm$ 1.73 \\
M11 &   $-0.121$  & 13.3  & $-0.11$ & 0.003 $\pm$ 1.61\\
\hline
\end{tabular}
\vspace{-0.15cm}
\label{table4}
\end{table}

\begin{figure*}
\begin{multicols}{2}
	\includegraphics[height=6cm,width=8.cm]{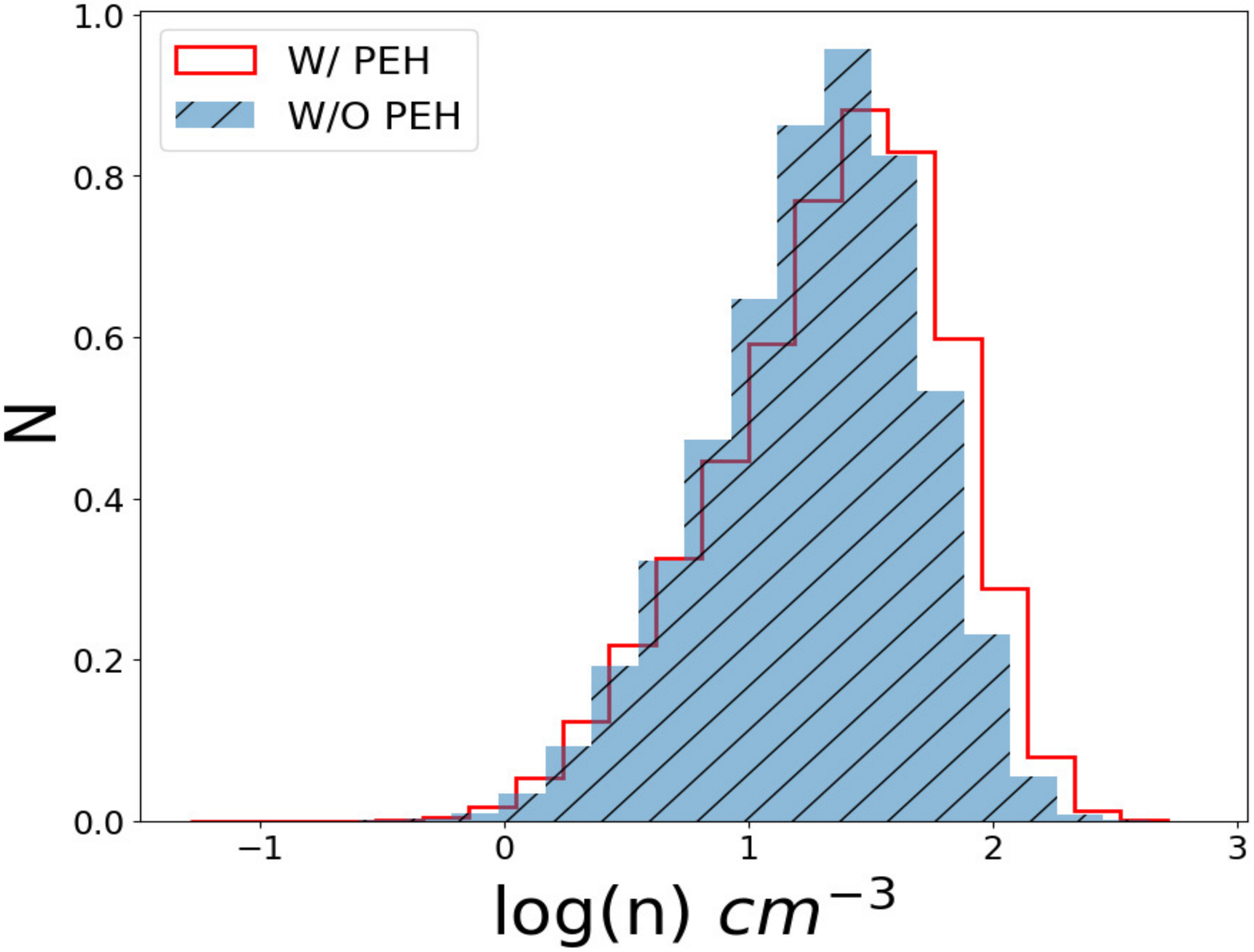}\par 
    \includegraphics[height=6cm,width=8.cm]{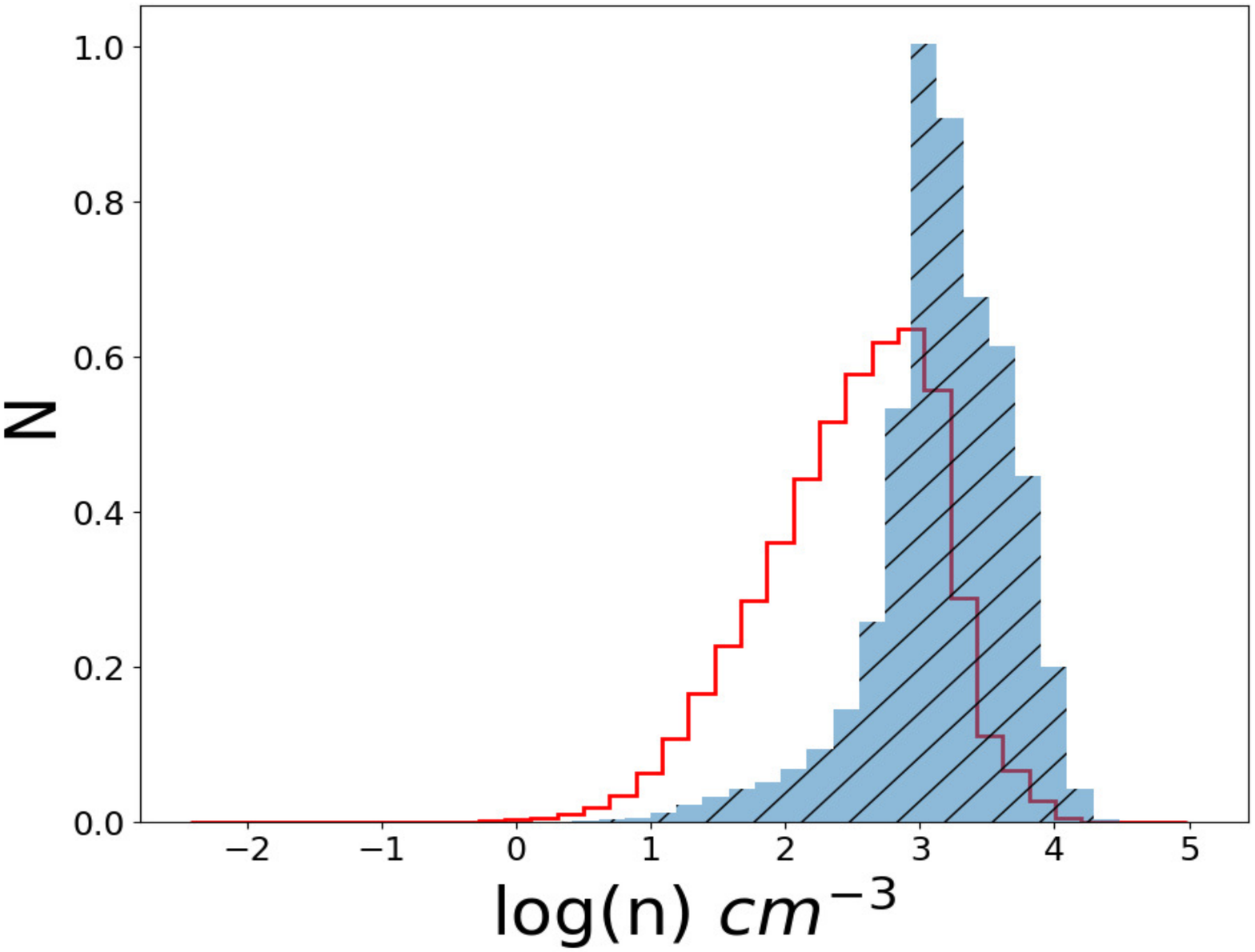}\par 
    \end{multicols}
    \vspace{-0.6cm}
\caption{Distributions of the density of the ISM ambient to SNe in two pairs of models M1 and M2 (left: as in Fig. \ref{fig}) and M3 and M4 (right: as in Fig. \ref{fig}).}
\vspace{-0.3cm}
\label{figfig}
\end{figure*}

To validate those arguments we investigated the ISM density ambient to SNe, metals and dust distributions in the halo, and  tracked halo gas particles back to the disk.  In tracking halo gas particles, we followed all the particles that are in the halo ($|z| > 2$ kpc) at 70 Myrs in both models in time up to 1 Gyr. Gas particles in the halo of M1 indeed tend to spend a longer time in the halo compared with the particles in M2 model (more than 70\% compared with slightly above 30\% of the particles did not make it back to the disk in the time scale of the simulation). The broader distribution of dust and metals along with the lower average metallicity and dust amount in the halo of model M1 compared with M2 supports our last argument. The investigation of the ISM density ambient to SNe, however, is not as expected where SNe tend to explode in slightly denser gas in M1 model compared with M2. This is probably because star formation has exhausted the dense gas in the model without PEH since in models with gas fraction 0.5 (M3 and M4) SNe indeed tend to explode in less dense gas in M3 compared with M4 (without PEH). This is hinted to in the phase diagrams in Fig. \ref{fig5} looking at the SNe shocked--heated gas. Fig. \ref{figfig} shows distributions of the density of the ISM ambient to SNe in the two pairs of models M1 and M2 (left: as in Fig. \ref{fig}) and M3 and M4 (right: as in Fig. \ref{fig})
Table 4 describes the change in the gas amount ($R_m$ ($M_{\sun}$)) in the halo ($|z| > 2$ kpc) over 1 Gyr in models with different gas fractions and PEH efficiencies. The choice of the 2 kpc cut off is somewhat arbitrarily and is mainly to ensure that we do not include gas in the thick disk in our calculation, especially for models with a high gas fraction. The errors associated with $R_m$ values account for the extent of the halo adopted (calculated for $|z| >$ 1, 2, and 3 kpc).  The extent of the disk is taken to be 24, 35, 40, and 50 kpc for models with f$_g$ = 0.03, 0.1, 0.3, and 0.5, respectively. The amount of gas in the halo increases with the gas fraction and PEH efficiency because of both influence SFRs in the models. $R_m$ is defined as follows:

\begin{equation}
R_m =  \begin{array}{rcl} M_{g,h,f} - M_{g,h,i} & M_{\sun} \end{array}
\end{equation}
where M$_{g, h, f}$ and M$_{g, h, i}$  are the total gas in the halo ($|z|$ $>$ 2 kpc) at T = 1 Gyr and T = 0 Gyr, respectively.

\section{Discussion}

\subsection{Star formation suppression}
Several processes lead to star formation suppression in galaxies such as stellar feedback that results in galactic winds (e.g. White \& Rees 1978; Dekel  \&  Silk  1986;  White  \&  Frenk 1991; Ceverino \& Klypin 2009; Muratov et al. 2015; El--Badry et al. 2016), AGN feedback (Nulsen et al. 2005; McNamara \& Nulsen 2007; Fabian 2012; Cicone et al. 2014), and environmental effects (e.g. tidal or ram--pressure stripping, Gunn \& Gott 1972; Moore et al. 1996; Abadi et al. 1999; Bekki 2014; Poggianti et al. 2017). Here, we have shown that photoelectric heating (PEH) of the gas by electrons ejected from dust grains suppresses star formation by reducing/increasing the density/temperature of the gas and probably creating an environment for SNe feedback to be more efficient. We tried to demonstrate the second point by investigating the gas amount in the halo, gas loading factors, time spent in the halo by gas particles, dust and metals distributions in the halo, and the ISM ambient to SNe in models with gas fractions 0.1 and 0.5. All these investigations, except for the ISM ambient to SNe in models with gas fraction 0.1 (see section 3.4 however), show that SNe efficiency is altered somehow in models with PEH. Models with PEH have a higher quantity of gas in their halos ($|z|$ $>$ 2 kpc), more than 70\% of the halo gas particles we tracked did not fall back to the disk in the time scale of the simulation, i.e. they are unavailable for star formation. Gas loading factors study also indicates the efficiency of outflows in affecting star formation.

Star formation occurs in models with different gas fractions, star formation recipes, dust models, and PEH efficiencies. The suppression on average ranges from negligible values to approximately a factor of five depending on the specific implementation. The parameter with the highest impact is the gas fraction because it controls the gas density which, in turn, controls star formation and photoelectric heating. We also note that the difference in the ISM properties between models with and without PEH becomes more considerable as the gas fraction increases. Accordingly, the SFR in models without PEH and with a higher gas fraction is enhanced compared to models with lower gas fraction, while in models with PEH and a higher gas fraction SFR is suppressed compared to models with a lower gas fraction.  The suppression lasts for a few Gyrs depending on the gas fraction (Butler et al. 2017), however its impact on the ISM may last for a longer time. It is worth noting that implementing a constant $F_e$ throughout the galaxy is an oversimplification which results in errors in estimating SFRs that are corresponding to standard deviation of 0.1 to 0.2 dex for models with gas fractions 0.1 and 0.5, respectively. Those errors become less significant compare to the suppression as the gas fraction increases.

\subsection{PEH effects on the ISM structure and abundance}

PEH and the subsequent reduction of the star formation activity influence the ISM in a non--linear manner where abundances, radial profiles, and 2D distributions of the different components of the ISM are altered. H$_2$ abundance is enhanced because of the lower consumption by star formation and dissociation by UV radiation from young stars. Moreover, the lower SNe rate enhances dust abundance by effectively reducing dust destruction. Thus, H$_2$ continues to form on dust grains, and dust grains continue to grow in molecular clouds. Most of the H$_2$ in those models, however, is not star--forming molecular gas (i.e. warm molecular hydrogen). On the contrary, metals are found in lower quantities in models with PEH, not only because of the lower amount of metals supplied by stars (lower SFRs, Hu et al. 2016) but also because of efficient dust growth which exhausted a high fraction of metals. Additionally, the inclusion of PEH leads to a flattening in the metallicity gradients.

PEH effects on the structure and properties of the ISM are found to be more significant as the gas fraction increases. In particular, models without PEH and with gas fraction of 0.5 developed pronounced gaseous and stellar clumps  (see Figs \ref{fig4} and \ref{fig7}). These clumps form gravitating systems towards which gas and disk stars gravitate, and they last for several dynamical times. Models with PEH formed less pronounced, short--lived ($<$ 100 Myrs) gaseous clumps along the spiral arms. These models can be linked to the clumpy disk galaxies at high and low redshifts (Cowie, Hu, \& Songaila 1995; van den Bergh et al. 1996; Conselice et al. 2004; Genzel et al. 2008;  Tadaki et al. 2014; Murata et al. 2014; Guo et al. 2015; Shibuya et al. 2016; Buck et al. 2017; Guo et al. 2018) where clumps in rotationally supported systems form via disk instabilities such as Toomre (1964) instability (Noguchi 1998; 1999; Shapiro et al. 2008; Dekel et al. 2009; Bournaud \& Elmegreen 2009) and spiral arms fragmentation (Inoue \& Yoshida 2018; 2019). The fate of these clumpy galaxies is still under debate, however, PEH could have had an important role in suppressing the formation, shortening the life time of these clumps, and further stabilize the disks that are thought to be the progenitors of present--day galaxies (e.g. Shlosman \& Noguchi 1993; Noguchi 1998; 1999; Hopkins, Quataert \& Murray 2012). 
 \begin{figure}
\includegraphics[height=6cm,width=8cm]{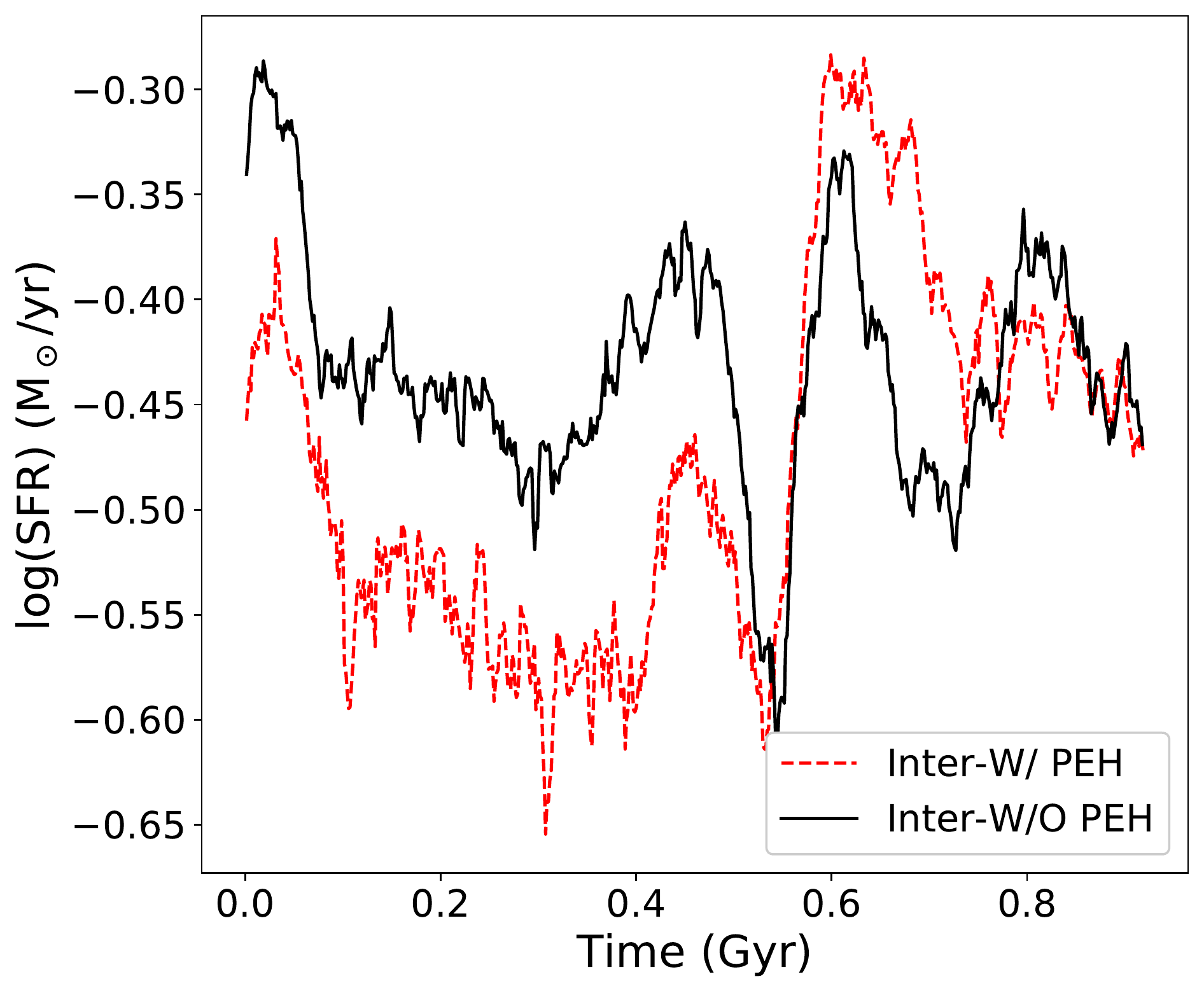}\par 
\caption{The time evolution of the SFR in models: M14 (red dashed, $F_e$ = 0.05) and M15 (black solid, $F_e$ = 0.0) of interacting pair of galaxies.}
\vspace{-0.3cm}
\label{fig9}
\end{figure}

Disk clumps are, arguably, able to migrate to the central regions of the galaxy and contribute to the bulge formation (Noguchi 1999; Immeli et al. 2004; Dekel et al. 2009). After evolving models M19 and M20 (Fig. \ref{fig7}) for 3 Gyrs, we report tentative evidence of bulge formation suppression via PEH. The model without PEH (M20) has a bulge one order of magnitude more massive than the bulge in the model with PEH (M19). Our forthcoming studies will investigate PEH effects on clumps formation and evolution.

\subsection{Effects of the physical models implemented}
\subsubsection{Star formation recipes and dust models}

When PEH is implemented in the models, H$_2$ density becomes sufficiently high for a H$_2$--dependent star formation recipe to maintain SFR similar to the SFR when a H--dependent recipe is implemented. This is not the case for models without PEH since their H$_2$ density is considerably low compared to the total gas density which results in suppression of the SFR. Hence, with PEH, implementing either of the star formation recipes makes little difference. The contrary occurs when a constant dust--to--gas ratio is implemented (e.g. Hu et al. 2017; Emerick et al. 2019) instead of implementing explicit dust evolution. SFRs are influenced in models with PEH and not in models without PEH. SFRs are overestimated in the case of models with PEH. Thus, the magnitude of SFR suppression is underestimated in this case, however, it could be also overestimated depending on the adopted dust--to--gas ratio.

\subsubsection{Galaxy mass}

Another important property that influences feedback processes in general and photoelectric heating in particular is the galaxy mass. Several studies showed that PEH suppresses star formation in dwarf galaxies (Forbes et al. 2016; Hu et al. 2017) and massive galaxies (Tasker 2011; Bekki 2015). Hence, we have run an extra set of models that includes Milky Way, M33, and SMC like (M$_h$ = 10$^{12}$, 10$^{11}$, and 10$^{10}$ M$_{\sun}$) with a gas resolution of about 2 $\times$ $10^5$ M$_{\sun}$. The models share the same parameters with the MW--like models presented here except for the mass, size, and resolution. We confirm the previous results and report that the magnitude of suppression depends on the gas fraction as well as the galaxy mass. The magnitude decreases as the mass decreases, however, it is not entirely clear when the gas fraction is small (f$_g$ $\sim$ 0.03).

\subsubsection{Interacting galaxies}

In Fig. \ref{fig9}, we explore whether or not our findings hold in the case of interacting galaxies. Models M14 (red dashed, $F_e$ = 0.05) and M15 (black solid, $F_e$ = 0.0) contain interacting pair of galaxies. At the beginning (in the first $\sim$ 500 Myrs) when the two galaxies are far apart, they act as isolated galaxies in terms of the PEH suppression of SFR, i.e. the model with PEH has SFR that is about 0.14 dex lower than the SFR in the model without PEH. At the time of their closest encounter ($\sim$ 600 Myrs), the model with PEH retained enough gas to starburst with higher magnitude/SFR and longer duration compared to the model without PEH. Afterwards, the SFR in the model with PEH drops slightly below the SFR in the model without PEH.

\section{Conclusions and summary} 
In this paper, we have used our original SPH code to study the influence of photoelectric heating (PEH) of the gas by electrons ejected from dust grains on the evolution of luminous disk galaxies. The evolution of the gas, dust, and ISRF are self--consistently implemented in the code. Dust evolution includes dust formation by stars, destruction by SNe, and growth in dense media. However, the code does not reach the necessary resolution to resolve the electron density in the ISM, hence, we treat the PEH efficiency ($F_e$) as a parameter that ranges between 0.05 and 0.003 (Bakes \& Tielens 1994). Our main results are summarized in the following.

(i) The diffuse heating caused by the PEH results in an ISM with an average temperature that is a factor of two to one order of magnitude higher compared to the ISM where PEH is switched off. This causes the density of the ISM to drop by a few and in turn, SFRs are suppressed. The suppression occurs in models with different gas fractions, star formation recipes, dust models, and PEH efficiencies. However, the magnitude of suppression (how much SFR is reduced when PEH is switched on) depends on the specific parameters adopted.

(ii) On the other hand, PEH enhances SNe feedback by altering the gas desnity. This enhancement increases the gas fraction in the halo. The gas driven out to the halo takes a long time (longer than the simulation timescale) to cool down and fall back onto the disk which makes it unavailable for star formation. The study of the gas mass loading factors in a few models indicates the efficiency of the outflows in affecting star formation.

(iii) The moderate consumption of the gas by star formation in models with PEH results in a higher abundance of all the different ISM components except for the gas--phase metals. Metals are less abundant in models with PEH compared to models without PEH because of their lower production rate by stars and their consumption by dust growth. Accordingly, PEH (through dust evolution) influences the global properties, radial profiles, and spatial distributions of the ISM. Perhaps one of the most important radial profiles are the metallicity profiles, we find that PEH flattens those profiles.

(v) Gas--rich disk galaxies in high redshift universe ($z$ $\sim$ 1--3) and the local universe (with less frequency) are found to have clumpy disks. Several mechanisms were proposed in the literature for clump formation, including violent disk instabilities and spiral arms fragmentation. When PEH is switched on, the violent disk instability is suppressed and less pronounced clumps are formed along the spiral arms. Furthermore, bulge formation via clump migration is also suppressed.

\section*{Acknowledgement}
OO is a recipient of an Australian Government Research Training Program (RTP) Scholarship. LC is a recipient of an Australian Research Council Future Fellowship (FT180100066) funded by the Australian Government.

\section*{Data availability}
The data underlying this article will be shared on a justified request to the corresponding author.
\section*{References}

\noindent Abadi M. G., Moore B., Bower R. G., 1999, MNRAS, 308, 947

\noindent Anderson M. E., Bregman J. N., 2010, ApJ, 714, 320

\noindent Aoyama S. et al., 2017, MNRAS, 466, 105

\noindent Asano R. S. et al., 2013, MNRAS, 432, 637

\noindent Bakes E. L. O., Tielens A. G. G. M., 1994, ApJ, 427, 822

\noindent Behroozi P. S., Conroy C.,  Wechsler R. H., 2010, ApJ, 717, 379

\noindent Beirao P. et al., 2012, ApJ, 751, 144

\noindent Bekki k., Shioya Y., 1998, ApJ, 497, 108

\noindent Bekki K., 2013, MNRAS, 432, 2298

\noindent Bekki K., Shigeyama T., Tsujimoto T., 2013, MNRAS, 428, L31

\noindent Bekki K., 2014, MNRAS, 438, 444

\noindent Bekki K., 2015, MNRAS, 449, 1625

\noindent Berentzen I., Shlosman I., Martinez-Valpuesta I., Heller C. H., 2007, ApJ, 666, 189

\noindent Bergin E. A. et al., 2004, ApJ, 612, 921

\noindent Bournaud F., Elmegreen B. G., 2009, ApJ, 694, L158

\noindent Bruzual G., Charlot S., 2003, MNRAS, 344, 1000

\noindent Buck T. et al., 2017, MNRAS, 468, 3628

\noindent Butler M. J. et al., 2017, 841, 82

\noindent Cacciato M., Dekel A., Genel S., 2012, MNRAS, 421, 818

\noindent Calzetti D. et al., 2000, ApJ, 533, 682

\noindent Capelo P. R. et al., 2018, MNRAS, 475, 3283

\noindent Cazaux S., Tielens A. G. G. M., 2004, ApJ, 604, 222

\noindent Ceverino D., Klypin A. 2009. ApJ, 695, 292

\noindent Chiang I-Da et al., 2018, ApJ, 865, 23

\noindent Choi E. et al., 2017, ApJ, 844, 31

\noindent Cicone C. et al., 2014, A\&A, 562, 25

\noindent Conselice C. J. et al., 2004, ApJ, 600, L139

\noindent Cowie L., Hu E., Songaila, A. 1995, AJ, 110, 1576

\noindent Cowie L. L., Songaila A., Hu E. M., Cohen J. G., 1996, AJ, 112, 839

\noindent Croxall K. V. et al., 2012, ApJ, 747, 81

\noindent Daddi E. et al., 2008, ApJ, 673, L21

\noindent Daddi E. et al., 2010, ApJ, 713, 686

\noindent Dekel A., Silk J., 1986, ApJ, 303, 39

\noindent Dekel A. et al., 2009, Nature, 457, 451

\noindent Draine B. T., 1978, ApJSS, 36, 595

\noindent Dwek E., Scalo J. M., 1980, ApJ, 239, 193

\noindent Dwek E., 1998, ApJ, 501, 643

\noindent El-Badry K. et al., 2016, ApJ, 820, 17

\noindent Elmegreen B. G., Bournaud F., Elmegreen D. M., 2008, ApJ, 688, 67

\noindent Emerick A., Bryan G. L., Mac Low Mordecai--Mark, 2019, MNRAS, 482, 1304

\noindent Fabian A. C., 2012, ARA\&A, 50, 455

\noindent Fisher D. B. et al., 2014, ApJ, 790, L30

\noindent Forbes J. C. et al., 2016, Nature, 535, 523

\noindent Forster Schreiber N. M. et al., 2009, ApJ, 706, 1364

\noindent Fukugita M., Hogan C. J., Peebles P. J. E., 1998, ApJ, 503, 518

\noindent Fukui Y., Kawamura A., 2010, ARA\&A, 48, 547

\noindent Gavazzi G., Pierini D., Boselli A., 1996, A\&A, 312, 397

\noindent Genzel R. et al., 2008, ApJ, 687, 59

\noindent Grossi M. et al., 2015, A\&A, 574, A126

\noindent Gunn J. E., Gott J. Richard III, 1972, ApJ, 176, 1

\noindent Guo Y. et al., 2015, ApJ, 800, 39

\noindent Guo Y. et al., 2018, ApJ, 853, 108

\noindent Habing H. J., 1968, Bull. Astron. Inst. Neth., 19, 421 

\noindent Herrera--Camus R. et al., 2018, ApJ, 861, 95 
 
\noindent Hill A. S., Mac Low Mordecai--Mark, Gatto A., Ibanez-Mejia J. C., 2018, ApJ, 862, 55

\noindent Hirashita H., 2000, PASJ, 52, 585

\noindent Hirashita H., 2013, POS (LCDU 2013) 027

\noindent Hitschfeld M., Kramer C., Schuster K. F., Garcia-Burillo S., Stutzki J., 2009, A\&A, 495, 795

\noindent Hopkins P. F., Quataert E., Murray N., 2012, MNRAS, 421, 3488

\noindent Hu Chia-Yu et al., 2016, MNRAS, 458, 3528

\noindent Hu Chia-Yu et al., 2017, MNRAS, 471, 2151

\noindent Hu Chia--Yu, 2019, MNRAS, 483, 3363

\noindent Immeli A., Samland M., Westera P., Gerhard O., 2004, ApJ, 611, 20

\noindent Ingalls J. G., Reach W. T., Bania T. M., 2002, 579, 289

\noindent Inoue A. K., 2011, Earth Planets and Space, 63, 1027

\noindent Inoue S., Yoshida N., 2018, MNRAS, 474, 3466

\noindent Inoue S., Yoshida N., 2019, MNRAS, 485, 3024

\noindent Jones A. P., 2000, JGR (Journal of Geophysical Research), 105, 10257

\noindent Komatsu E.  et al., 2011, ApJS, 192, 18

\noindent Kormendy J., 2013, In XXIII Canary Islands Winter School of Astrophysics, Secular Evolution of Galaxies eds. J. Falcon-Barroso \& J. H. Knapen (Cambridge: Cambridge University
Press), p. 1

\noindent Kormendy J., Kennicutt R. C., 2004, ARAA, 42, 603

\noindent McNamara B. R., Nulsen P. E. J., 2007, ARA\&A, 45, 117

\noindent Mo H. J. et al., 2005, MNRAS, 363, 1155

\noindent Moore B. et al., 1996, Nature, 379, 613 

\noindent Moster B. P., Naab T., White S. D. M., 2013, MNRAS, 428, 3121

\noindent Murata K. L. et al., 2014, ApJ, 786, 15

\noindent Muratov A. L. et al., 2015, MNRAS, 454, 2691

\noindent Naab T., Ostriker J. P., 2017, ARA\&A, 55, 59

\noindent Navarro J. F., Frenk C. S., White S. D. M., 1996, ApJ, 462, 563

\noindent Noguchi M., 1998, Nature, 392, 253

\noindent Noguchi M., 1999, ApJ, 514, 77

\noindent Norman C. A., Sellwood J. A., Hasan H., 1996, ApJ, 462, 114

\noindent Nozawa T., Kozasa T.,  Habe A., 2006, ApJ, 648, 435

\noindent Nulsen, P. E. J. et al., 2005, ApJ, 628, 629

\noindent Okada Y. et al., 2013, A\&A, 553, A2

\noindent Osman O., Bekki K., Cortese L., 2020, MNRAS, submitted

\noindent Ostriker J. P., Peebles P. J. E., 1973, ApJ, 186, 467

\noindent Pfenniger D., Norman C., 1990, ApJ, 363, 391

\noindent Poggianti B. M. et al., 2017, ApJ, 844, 21

\noindent Puech M., 2010, MNRAS, 406, 535

\noindent Remy--Ruyer A. et al., 2014, A\&A, 563, A31

\noindent Ribeiro B. et al., 2017, A\&A, 608, A16

\noindent Robertson B. E., Bullock J. S., 2008, ApJ, 685, L27

\noindent Rollig M. et al., 2006, A\&A, 451, 917

\noindent Rosen A., Bregman J. N., 1995, ApJ, 440, 634

\noindent Savage B. D., Sembach K. R., 1996, AR A\&A, 34, 279, 329

\noindent Schneider R., Ferrara A., Natarajan P., Omukai K., 2002, ApJ, 571, 30

\noindent Schneider R., Ferrara A., Salvaterra R., Omukai K., Bromm V., 2003, Nature, 422, 869

\noindent Shapiro K. L. et al., 2008, ApJ, 682, 231

\noindent Shibuya T., Ouchi M., Kubo M., Harikane Y., 2016, ApJ, 821, 72

\noindent Shlosman I., Noguchi M., 1993, ApJ, 414, 474

\noindent Sofia U. J., 2004,  ASP, 309, 393

\noindent Storzer H., Stutzki J.,  Sternberg A., 1996, A\&A, 310, 592

\noindent Sutherland R. S., Dopita M. A., 1993, ApJS, 88, 253

\noindent Tacconi L. J. et al., 2008, ApJ, 680, 246

\noindent Tadaki Ken-ichi et al., 2014, ApJ, 780, 77

\noindent Tasker E. J., 2011, ApJ, 730, 11

\noindent Thornton K., Gaudlitz M., Janka H.-Th., Steinmetz M., 1998, ApJ, 500, 95

\noindent Toomre A., 1964, ApJ, 139, 1217

\noindent van den Bergh S. et al., 1996, AJ, 112, 2

\noindent Veilleux S., Cecil G., Bland--Hawthorn J., 2005, ARA\&A, 43, 769

\noindent Wakelam V. et al., 2017, Molecular Astrophysics, 9, 1

\noindent Watson W. D., 1972, ApJ, 176, 103

\noindent Weiner B. J. et al., 2006, ApJ, 653, 1027

\noindent Weingartner J. C., Draine B. T., 2001, ApJSS, 134, 263

\noindent White S. D. M., Rees M. J., 1978, MNRAS, 183, 341

\noindent White S. D. M., Frenk C. S., 1991, ApJ, 379, 52

\noindent Wolfire M. G. et al., 1995, ApJ, 443, 152

\noindent Wolfire M. G., Mckee C. F., Tielens A. G. G. M., 2003, ApJ, 587, 278

\noindent Zhukovska S., 2014, A\&A, 562, A76

\end{document}